\documentclass[aps,pra,twocolumn,groupedaddress,longbibliography]{revtex4-1}

\usepackage{amsmath}
\usepackage{graphicx}
\usepackage{siunitx}
\usepackage[american]{circuitikz}
\usepackage{physics}

\usepackage{color}
\usepackage{url}

\usepackage[colorlinks]{hyperref}
\hypersetup{%
       plainpages=true,
       breaklinks=true,
       hypertexnames=false,
       pageanchor=true,
       colorlinks=true,
       linkcolor=[rgb]{0,0.22,0.43},
       citecolor=[rgb]{0,0.22,0.43},
       urlcolor=[rgb]{0,0.22,0.43},
       anchorcolor={black}
     }

\begin{document}


\title{Cavity-free vacuum-Rabi splitting in circuit quantum acoustodynamics}


\author{Andreas Ask}
\affiliation{Department of Microtechnology and Nanoscience (MC2), Chalmers University of Technology, SE-41296 G\"{o}teborg, Sweden }

\author{Maria Ekstr\"{o}m}
\affiliation{Department of Microtechnology and Nanoscience (MC2), Chalmers University of Technology, SE-41296 G\"{o}teborg, Sweden }

\author{Per Delsing}
\affiliation{Department of Microtechnology and Nanoscience (MC2), Chalmers University of Technology, SE-41296 G\"{o}teborg, Sweden }

\author{G\"{o}ran Johansson}
\affiliation{Department of Microtechnology and Nanoscience (MC2), Chalmers University of Technology, SE-41296 G\"{o}teborg, Sweden }


\date{\today}

\begin{abstract}
Artificial atoms coupled to surface acoustic waves (SAWs) have played a crucial role in the recent development of circuit quantum acoustodynamics (cQAD). In this paper, we have investigated the interaction of an artificial atom and SAWs beyond the weak coupling regime, focusing on the role of the interdigital transducer (IDT) that enables the coupling. We find a parameter regime in which the IDT acts as a cavity for the atom, rather than an antenna. In other words, the atom forms its own cavity. Similar to an atom coupled to an explicit cavity, this regime is characterized by vacuum-Rabi splitting, as the atom hybridizes with the phononic vacuum inside the IDT. This hybridization is possible because of the interdigitated coupling, which has a large spatial extension, and the slow propagation speed of SAWs. We work out a criterion for entering this regime from a model based on standard circuit-quantization techniques, taking only material parameters as inputs. Most notably, we find this regime hard to avoid for an atom on top of a strong piezoelectric material, such as LiNbO$_3$. The SAW-coupled atom on top of LiNbO$_3$ can thus be regarded as an atom-cavity-bath system. On weaker piezoelectric materials, the number of IDT electrodes need to be large in order to reach this regime. 
\end{abstract}


\maketitle

\section{Introduction}

In the last two decades, circuit quantum electrodynamics (cQED) has emerged as one of the premier platforms for the study of light-matter interactions \cite{Blais2004,Wallraff2004,Gu2017,Wendin2017}. At the heart of its success lies the strong nonlinearity provided by the Josephson junction, enabling  engineering of artificial atoms and other novel non-linear quantum devices. Recently, there has been rapid experimental progress in coupling the artificial atom to a variety of acoustic waves in solids, such as: propagating surface acoustic waves (SAWs)~\cite{Gustafsson2014}, SAW resonators \cite{Manenti2017,Moores2017,Noguchi2017,Bolgar2018,Satzinger2018}, and bulk acoustic wave resonators \cite{Chu2017,Kervinen2018}. These hybrid systems sparked the recent development of circuit quantum acoustodynamics (cQAD), where acoustic phonons take over many of the roles played by photons in cQED. Furthermore, SAWs in the quantum regime have played important roles in a number of recent theoretical proposals, including: a general transducer between hybrid quantum systems~\cite{Schuetz2015}, a microwave to optical frequency conversion protocol \cite{Shumeiko2016}, and could provide a platform for general quantum computation via delayed feedback, utilizing the slow propagation speed of SAWs ($\approx 3000$ m/s) \cite{Pichler2017a}. 

SAW devices have been successfully integrated with classical electronic circuits since the invention of the interdigital transducer (IDT) \cite{Datta} in the mid 1960s. Their primary use has been as bandpass filters and delay lines. With the advent of SAWs to the quantum regime, it is possible that also quantum technologies, based on quantized electrical circuits, can take advantage of the unique properties of SAW. Of special interest is the superconducting artificial atom, since it constitutes the main ingredient in many implementations of quantum computation. So far, the type of superconducting atom that has been successfully coupled to SAW is the transmon \cite{Koch2007}, which is also the atom studied in our work. 



The coupling between a transmon and SAWs is enabled by forming the large shunt capacitance of the atom into an IDT \cite{Gustafsson2014}, which serves as a phononic antenna. The conversion to phonons through this antenna relies on the piezoelectric effect, and relaxation into other decay channels can be made small \cite{Gustafsson2014}. Since the spatial extension of the IDT exceeds many wavelengths in space, the coupling is not pointlike, as it normally is for both natural atoms at optical frequencies, and artificial atoms coupled to microwave photons. Earlier theoretical studies accounted for the interdigitated coupling by modeling an atom coupled to a bosonic bath at multiple points in space \cite{FriskKockum2014,Kockum2018a}. The coupling points were considered pointlike, and a standard master equation derivation for the atom was performed, in which the bath was traced out while assuming weak system-bath coupling. Under this assumption, the atom was predicted to have a frequency-dependent relaxation rate. This prediction was also recently verified experimentally \cite{Moores2017}. 
In fact, most recent experiments in cQAD involves an atom on top of a weakly piezoelectric material such as quartz \cite{Manenti2017,Noguchi2017} or gallium arsenide (GaAs) \cite{Gustafsson2014}, in which the weak coupling approximation holds. However, it may be advantageous to use a strongly piezoelectric material, such as lithium niobate (LiNbO$_3$), in order to increase the coupling strength between the atom and the substrate. Additionally, any auxiliary IDTs used to send or receive SAWs would benefit from the increased conversion efficiency \cite{Ekstrom2017}. Under the assumption that the IDTs should be impedance matched (typically to 50 $\Omega$), they would also have a larger bandwidth on a stronger piezoelectric. 

Motivated by these possible advantages, we have investigated the interaction between an artificial atom and SAWs for varying degrees of coupling strength, starting from a model based on a quantized electrical circuit. \cite{Yurke1984,Devoret1995}. Interestingly, as the coupling strength is increased, we find a regime in which the coupling-IDT acts as a cavity for the atom, rather than an antenna. In other words, since the IDT constitutes a major part of the atom itself, the atom forms its own cavity. In contrast to earlier studies \cite{FriskKockum2014}, our model is built on design and material parameters of the IDT and the piezoelectric substrate. We can thus predict the specific parameter regime in which the IDT forms a cavity. We will name the two possible regimes the \textit{antenna}- and \textit{cavity regime} respectively, referring to the role of the IDT. Until now, it is the antenna regime that has been realized on weak piezoelectrics such as GaAs \cite{Gustafsson2014}.  

The existence of a cavity regime is expected due to the spatial extension of the IDT. If the coupling strength becomes strong enough, the atom has time to interchange an excitation with the part of the substrate covered by the IDT multiple times before it leaves the system at either end. Such coherent exchange of energy leads to vacuum-Rabi splitting as the atom hybridizes with the phononic vacuum inside the IDT. More precisely, we find that vacuum-Rabi splitting occurs for a SAW-coupled atom as long as its decay rate exceeds the inverse phononic traveling time across the IDT, $\gamma T \geq 1$. 

We additionally find that the cavity regime is unavoidable for an atom on top of LiNbO$_3$, with a reasonable number of metal electrodes (fingers) in the IDT ($> 4 $ single finger pairs). Conversely, an atom on GaAs never reaches this regime unless the number of finger pairs approaches $ 30$, which would make the anharmonicity of the atom very small by decreasing its charging energy significantly \cite{Koch2007}. Realizing the true role of the IDT in the particular parameter regime of interest is important as the field of cQAD evolves and effective models of more elaborate set-ups are required.

This article is organized as follows. We map the the SAW-coupled atom onto an electrical circuit in section \ref{ingredients}. From that we derive time delay differential equations describing the motion of the atom operators. We solve the linearized equations of motion in section \ref{results}, relevant in the single excitation regime of a weakly driven atom for arbitrary coupling strength, and calculate the charge response of the atom, as well as an acoustic and electric reflection coefficient. The addition of an electric gate becomes important in the cavity regime as it allows for direct probing of the system dynamics by bypassing the IDT. Finally, we conclude the paper in section \ref{conclusion}.

\begin{figure*}
\includegraphics[width=\textwidth]{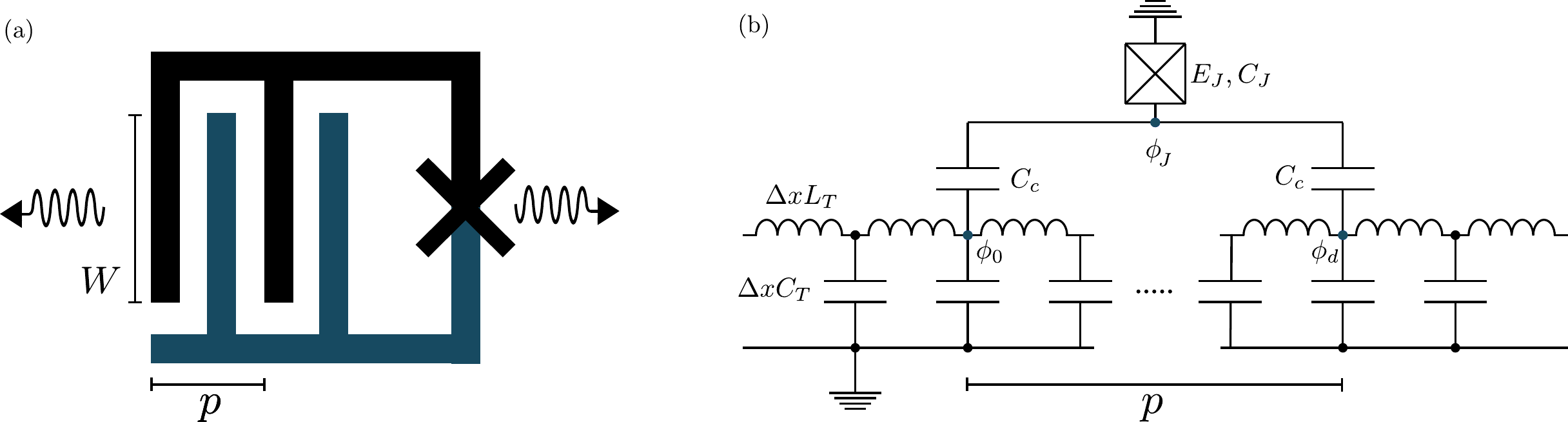}
\caption{\label{idt_circuit}(a) Cartoon of an IDT shunted by a Josephson junction. The distance between two fingers of equal voltage is called the finger pitch $p$ and defines the resonance frequency of the IDT $\omega_{\text{IDT}}=2\pi v_s/p$. The overlap of the fingers are denoted $W$ and sets the width of the waves emitted by the IDT. (b) A discrete circuit representation of the schematic model in (a). A Josephson junction is  capacitively coupled to an acoustic transmission line at two points in space, separated a distance equal to the finger pitch of the IDT. The corresponding circuit Lagrangian can be seen in Eq.~\eqref{circuit_lagrangian}.}
\end{figure*}

\section{Circuit model}
\label{ingredients}
In this section we build our circuit model of a SAW-coupled atom, using standard circuit quantization techniques \cite{Yurke1984,Devoret1995}.

The model we have in mind is a Josephson junction that couples capacitively to a transmission line at multiple points in space. Each finger pair of the IDT corresponds to one capacitive coupling point between the atom and the transmission line (see Fig.~\ref{idt_circuit}). 
Our model has two advantages: it is a straightforward circuit representation of the phenomenological model previously used to model the SAW-coupled atom \cite{FriskKockum2014}, and can thus be used to bridge between that model and the classical lumped element model used to interpret several experiments \cite{Gustafsson2014,Moores2017}. Additionally, it captures the effects of the IDT in a simple and intuitive way. 

For transparency, we only consider two coupling points (two finger pairs) initially, but generalize all final results to an arbitrary number of coupling points. The final model describes the interaction between an artificial transmon atom and a 1-D SAW transmission line, enabled by an IDT with arbitrary number of fingers. 

Before we construct our model, we review the classical literature in order to map important SAW parameters onto lumped circuit elements, and make our work self-contained. We follow closely the description in \cite{Datta}, to which we refer the reader for a thorough introduction to SAWs and IDTs.

\subsection{Freely propagating SAWs}
Surface acoustic waves are Rayleigh waves propagating on the surface of materials, and extends approximately one wavelength into the bulk \cite{Stoneley1955}. When SAWs propagates on top of a piezoelectric material there is also an electric potential, and associated electric field, propagating together with the mechanical wave. Its motion is primarily determined by the solutions to the acoustic wave equations, since a negligible amount of energy resides in the electric field \cite{Simon1996}. However, in a transmission-line model of SAWs, the mechanical degrees of freedom are neglected by choosing a characteristic impedance so that all energy is accounted for by the electric field alone.  

In order to describe SAWs in a transmission-line picture, we introduce a phenomenological current $I$, which should not be mistaken as a physical current since there is no net charge traversing the surface. From this current we can define the time-averaged power flow of a wave $P = V_s I/2$,
where $V_s$ is the voltage of the SAW potential. The characteristic impedance of a SAW transmission line then becomes $Z_0 = V_s^2/2P$.
Knowing the characteristic impedance, and the wave velocity $v_s$, we can calculate the inductance per unit length $L_T= Z_0/v_s$, and the capacitance per unit length $C_T = 1/Z_0v_s$.
This allows us to write down the time-averaged energy stored per unit length $U = C_T V_s^2/2$.
We note that, due to how the power flow was defined, this is the total energy stored in the wave, containing both the electric and acoustic field contributions. Thus, we have the freedom to describe SAWs in terms of its electric properties alone.

The impedance $Z_0$ is not only a material property since the total power depends on the width of the SAW wave. For waves emitted by an IDT, this is the overlapping part of the IDT fingers, denoted $W$ in Fig.~\ref{idt_circuit}(a). It is therefore common practice to define an impedance $z_0$ and corresponding admittance $y_0 = z_0^{-1}$ independent of the beam width \cite{Datta},
\begin{equation}
Z_0 = \frac{z_0\lambda}{W} = \frac{\lambda}{y_0W},
\end{equation}
where $\lambda$ is the wavelength of the SAW. The admittance $y_0$ is a material parameter that can be found in tables (see e.g. \citep{Datta}) in order to calculate $Z_0$. However, it is also given in terms of other material parameters according to the relation $K^2y_0 = 2 \pi \epsilon_{\infty}v_s$ (see \citep{Datta} for a derivation of this equation), where $K^2$ is the piezoelectric coupling coefficient, often used to quantify the degree of piezoelectricity in a material, and where $\epsilon_{\infty} = \epsilon_p + \epsilon_0$ is an effective permittivity given by the sum of the permittivity of the substrate material $\epsilon_p$ and the permittivity of vacuum $\epsilon_0$. Thus, the impedance can be expressed as,
\begin{equation}
Z_0 = \frac{K^2}{W\epsilon_{\infty} \omega}.
\label{imp}
\end{equation}

In contrast to microwave photons, freely propagating SAWs in a one-dimensional model are not dispersion free, which is manifested in the frequency dependence of Eq.~\eqref{imp}. It originates from the fact that SAWs are not isotropic in the direction perpendicular to its direction of propagation due to the finite length of the IDT fingers. For systems with narrow bandwidth and sharp resonance peaks, like atoms on GaAs, this frequency dependence is slow and can be neglected. 

With the characteristic impedance in place we have everything we need in order to describe the freely propagating SAWs in terms of the distributed element circuit in Fig.~\ref{idt_circuit}(b). 
\label{z0}

\subsection{Coupling capacitance}
\label{cc}
The coupling capacitance serves two purposes: it couples the atom to SAWs, and it forms the IDT capacitance. The IDT capacitance also constitutes the big shunt capacitance of the transmon. Since our model relies on a single parameter to account for both of these effects, we could in principle choose which effect we derive our coupling capacitance from. We will take a simple approach where we distribute the total capacitance of the IDT equally among the finger pairs of the IDT, the resulting capacitance for one finger pair is then used as the coupling capacitance. It turns out that this rather simple approach yields approximate results in good agreement with what is expected from earlier studies \cite{Gustafsson2014}. 

The total capacitance of an IDT depends on its geometry, and for arbitrary shapes the capacitance is most commonly calculated using commercial software. For periodic IDTs with metalization ratio $1/2$, such as the one in Fig.~\ref{idt_circuit}(a), the capacitance is given by the simple expression $C_{\Sigma} = n\epsilon_{\infty}W$, where $n$ is the number of finger pairs. The coupling capacitance of our model then becomes,
\begin{equation}
C_c = \epsilon_{\infty}W.
\end{equation} 
It is common in many application to use a double finger structure in order to mitigate internal mechanical reflections between the fingers. The capacitance is then reduced by a factor of $\sqrt{2}$. In this work, we focus on single finger IDTs, however, all results can be altered to also account for double finger IDTs by simply reducing the coupling capacitance to $C_c = \epsilon_{\infty}W/\sqrt{2}$. For a detailed derivation of the IDT capacitance we refer the reader to \cite{Morgan2007}.

\subsection{Equations of motion for the atomic operators}
\label{model}


Starting from the circuit in Fig.~\ref{idt_circuit}(b), we obtain the following Lagrangian, given in the node flux representation \cite{Devoret1995},
\begin{equation}
\begin{split}
L &= 
\frac{2C_c + C}{2} \dot{\phi_J}^2 + \frac{C_c}{2} \dot{\phi_0}^2 + \frac{C_c}{2} \dot{\phi_d}^2 - C_c\dot{\phi_0}\dot{\phi_J} \\
& - C_c\dot{\phi_d}\dot{\phi_J} + \sum_{i=-\infty }^{\infty} \frac{\Delta x C_T }{2}\dot{\phi_i}^2 \\
& - E_J \cos\left ( 2 \pi \phi_J/\Phi_0\right ) - \sum_{i=-\infty }^{\infty} \frac{(\phi_{i+1} - \phi_{i})^2}{2\Delta x L_T},
\label{circuit_lagrangian}
\end{split}
\end{equation}
where the flux is defined as the time integral of the node voltage $\phi(t) = \int_{-\infty}^tV(t')dt'$.
The two coupling points are denoted $0$ and $d$ and are separated a distance $d\Delta x$ apart, corresponding to the finger pitch $p$ of an IDT, see Fig.~\ref{idt_circuit}(a). The flux at the two coupling points and at the Josephson junction defines what we refer to as the system variables. 

The conjugate momentum to the flux is defined as $p_i = \partial L/\partial \dot{\phi}_i$ and has the unit of charge. We obtain a Hamiltonian from the Legendre transformation of $L$,
\begin{equation}
\begin{split}
H & = \frac{p_J^2}{2C_J} + \frac{p_0^2}{2}\left (\frac{1}{C_J} + \frac{1}{C_c} \right ) +  \frac{p_d^2}{2}\left (\frac{1}{C_J} + \frac{1}{C_c} \right ) + \frac{p_0p_J}{C_J}  \\ & + \frac{p_dp_J}{C_J}  + \frac{p_0p_d}{C_J}  
 +  E_J \cos (2 \pi \phi_J/ \phi_0) \\
 & + \sum_{\substack{i = -\infty}}^{\infty}\left ( \frac{p_i^2}{2\Delta x C_T} + \frac{(\phi_{i+1} - \phi_{i})^2}{2\Delta x L_T} \right ).
 \label{hamiltonian}
 \end{split}
 \end{equation}
We quantize our variables in the canonical way, promoting both the system and transmission line variables to operators and imposing commutation relations,
\begin{align}
[\hat{\phi}_i,\hat{p}_j] &= i\hbar\delta_{ij}, \\
[\hat{\phi}_i,\hat{\phi}_j] &= [\hat{p}_i,\hat{p}_j] = 0.
\end{align}
We also set $\hbar = 1$ and drop the hat on the operators from now on. Next, we calculate the equations of motion (EOM) for the $\phi$ and $p$ operators, while going to the continuum limit $\Delta x \rightarrow 0$. In the continuum limit, the node flux takes the value of the flux field evaluated at the position of the node, $\phi_i(t) = \phi(x_i,t)$.   From the EOM for the transmission line operators,  we can derive the Klein-Gordon equation for a massless scalar field, describing the motion of the freely propagating field \cite{Peropadre2013},
\begin{equation}
\frac{\partial^2\phi(x,t)}{\partial t^2	} - \frac{1}{L_TC_T}\frac{\partial^2 \phi(x,t)}{\partial x^2} = 0.
\label{wave}
\end{equation}
The general solution to Eq.~\eqref{wave} can be divided into left and right propagating fields,
\begin{equation}
\phi(x,t) = \phi\left(x - v_st\right) + \phi \left(x + v_st \right),
\end{equation}
with the SAW velocity $v_s = 1/\sqrt{L_TC_T}$. We will denote the direction of propagation by an arrow:
\begin{equation}
\overrightarrow{\phi}(x) = \phi(x-v_st).
\end{equation}
For the system operators we get the following EOM,
\begin{align}
\dot p_J &= -E_J\left(\frac{2 \pi}{\phi_0}\right) \sin(2 \pi \phi_J/\phi_0), \label{pj1} \\
\dot p_0 & = -\sqrt{\frac{C_T}{L_T}} \left( \dot \phi_0 -  \dot \phi_0^{\text{in}} \right ) = -\sqrt{\frac{C_T}{L_T}} \left( -\dot \phi_0 +  \dot \phi_0^{\text{out}} \right ), \label{in1} \\
\dot p_d &= -\sqrt{\frac{C_T}{L_T}} \left( \dot \phi_d -  \dot  \phi_d^{\text{in}} \right ) = -\sqrt{\frac{C_T}{L_T}} \left( -\dot \phi_d + \dot  \phi_d^{\text{out}} \right ), \label{in2} \\
 \dot \phi_J &= \frac{p_J}{C_J} + \frac{p_0}{C_J} + \frac{p_d}{C_J}, \\
\dot \phi_0	& = \left (\frac{1}{C_J} + \frac{1}{C_c}\right ) p_0+ \frac{p_J}{C_J} + \frac{p_d}{C_J}, \\
 \dot \phi_d & = \left (\frac{1}{C_J} + \frac{1}{C_c}\right )p_d + \frac{p_J}{C_J} + \frac{p_0}{C_J}.
\label{eomSys}
\end{align} 
The in- and out-fields in Eqs. \eqref{in1} and \eqref{in2} are referring to the field components that propagates into, or away from, each coupling point,
\begin{align}
\phi_0^{\text{in}} & =  \overrightarrow\phi(0^-) + \overleftarrow\phi(0^+), \\
\phi_0^{\text{out}} & =   \overleftarrow\phi(0^-) + \overrightarrow\phi(0^+), 
\end{align}
and similarly for the second coupling point. The sign refers to an infinitesimal displacement on either side, e.g. $\phi(0^-) = \phi(0 - \Delta x)$. In deriving the Eqs. \eqref{pj1}-\eqref{eomSys} we also used the boundary conditions that the field has to be continuous over the coupling points,
\begin{align}
\phi_0 &= \overrightarrow{\phi}(0^-) + \overleftarrow{\phi}(0^-) = \overrightarrow{\phi}(0^+) + \overleftarrow{\phi}(0^+),\\
\phi_d &= \overrightarrow{\phi}(d^-) + \overleftarrow{\phi}(d^-) = \overrightarrow{\phi}(d^+) + \overleftarrow{\phi}(d^+).
\end{align}

We can simplify the Eqs. \eqref{pj1}-\eqref{eomSys} further and write the EOM in terms of the transmon operators $\phi_J$ and $p_J$ alone, together with input-output relations for the fields propagating away from the two ends of the IDT,
\begin{align}
\dot \phi_J &=  \frac{p_J}{2C_c + C_J} + \frac{C_c}{2C_c+C_J}\partial_t \bigg(2 Z_0 \frac{C_c}{2C_c+C_J}\big(p_J \nonumber\\
& + p_J(t-\tau)\big)  + \Big(\overrightarrow\phi(0^-) +\overrightarrow\phi(0^-,t-\tau)  \label{eom1} \\
  &+ \overleftarrow\phi(d^+) + \overleftarrow\phi(d^+,t-\tau)\Big) \bigg),\notag \\
\dot p_J &=  -E_J\left(\frac{2 \pi}{\phi_0}\right) \sin(2 \pi \phi_J/\phi_0),\label{eom2} \\
\overleftarrow{\phi}(0^-) &= \overleftarrow{\phi}(d^+,t-\tau) + \frac{Z_0C_c}{2\left(2C_c + C_J\right)}\left (p_J + p_J(t-\tau)\right),\label{output1} \\
\overrightarrow{\phi}(d^+) & = \overrightarrow{\phi}(0^-,t-\tau) + \frac{Z_0C_c}{2\left(2C_c + C_J\right)}\left( p_J + p_J(t-\tau)\right ) \label{output2}.
\end{align}
Note that we have left out the time dependence of the operators that are not time-shifted. In order to derive Eqs. \eqref{eom1}-\eqref{output2}, we used that the in- and outgoing field between the two coupling points are related by a time delay,
\begin{align}
\overrightarrow{\phi}(d^-,t) &= \overrightarrow{\phi}(0^+,t-\tau),
\label{taushiftright}\\
\overleftarrow{\phi}(0^+,t) &= \overleftarrow{\phi}(d^-,t-\tau),
\label{taushiftleft}
\end{align}
where $\tau = p/v_s$ is the time of flight between the two coupling points, set by their separation $p$ (the finger pitch) and the speed of SAWs $v_s$. From $\tau$ we can define a characteristic resonance frequency of the system $\omega_{\text{IDT}} \equiv 2 \pi /\tau$, i.e. the center frequency of the IDT, which is the frequency at which waves interfere constructively at the coupling points. Maximum conversion to phonons occur when the atom is on resonance with the IDT center frequency, and will be the main regime of interest in this work. 

The system of equations in \eqref{eom1}-\eqref{output2} is a set of non-linear time delay differential equations, which are generally hard to solve. Time delays in quantum systems have been addressed recently in several studies, e.g. \cite{Grimsmo2015,Pichler2016}, all relying on the rotating wave approximation (RWA). The RWA is only valid for weakly coupled systems. We are interested in a wide range of coupling strengths, all the way up to the so called ultrastrong coupling regime (USC) \cite{Forn-Diaz2018,Kockum2018}. 
To our knowledge, there is currently no method that combines USC and time delays in a satisfactory way. Recent methods using matrix product states \cite{Sanchez-Burillo2014} might not be suitable because of the large spatial extension of the IDT. 
We therefore linearize the Josephson current in Eq.~\eqref{eom2} and solve the EOM exactly in the frequency domain, valid in the single excitation regime for arbitrary coupling strengths. For a weakly anharmonic system, like the transmon, the Josephson non-linearity can be added as a perturbation once the harmonic solution has been obtained, in the spirit of black box quantization \cite{Nigg2012}, but that is left for future work.

\subsection{Linearized EOM for an arbitrary number of coupling points}
\begin{figure}
\includegraphics{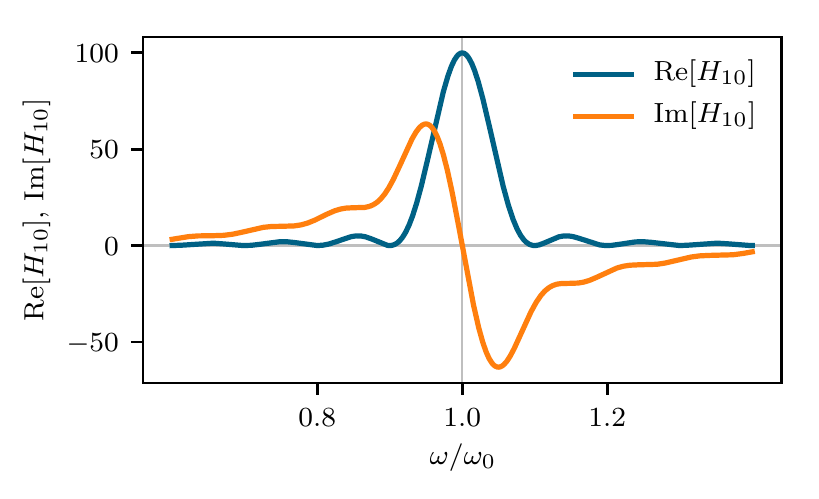}
\caption{\label{GandB}The real (blue) and imaginary (orange) part of $H_n$, defined in Eq.~\eqref{H} for $n=10.$ The real and imaginary parts are related by a Hilbert transform according to the Kramers-Kronig relation. The real part is proportional to the atomic decay rate, which makes it frequency-dependent, the imaginary part is responsible for the vacuum-Rabi splitting discussed in section \ref{results}.}
\end{figure}

To study the linear response of our system we expand the Josephson current in Eq.~\eqref{eom2} and disregard all but the linear contribution, from which we define the Josephson inductance $L_J = \Phi_0^2/E_J4\pi^2$. In the frequency domain Eq.~\eqref{eom1}-\eqref{output2} then becomes, 
\begin{align}
\phi_J & = -\frac{ip_J}{\omega(2C_c + C_J)} +\frac{C^2_cZ_0p_J(2 + 2e^{-i \omega \tau})}{2(2C_c + C_J)^2} \\
 & + \frac{C_c}{2C_c+C_J} \left(\phi^{\text{in}}_0 + \phi^{\text{in}}_d \right)(1 + e^{-i \omega \tau}), \nonumber \\
 p_J & = \frac{i\phi_J}{\omega L_J} ,\\
\phi_0^{\text{out}} &= \phi^{\text{in}}_de^{-i \omega \tau}  + \frac{Z_0C_c p_J(1 + e^{-i \omega \tau})}{2(2C_c + C_J)}, \\
 \phi^{\text{out}}_d& = \phi^{\text{in}}_0e^{-i \omega \tau} + \frac{Z_0C_c p_J(1 + e^{-i \omega \tau})}{2(2C_c + C_J)},
\end{align}
where $\phi^{\text{in}}_0 = \overrightarrow{\phi}(0^-)$ and $\phi^{\text{in}}_d = \overleftarrow{\phi}(d^+)$ refers to the field components that propagates into the IDT at either end, and $\phi^{\text{out}}_0 = \overleftarrow{\phi}(0^-)$ and $\phi^{\text{out}}_d = \overrightarrow{\phi}(d^+)$ refers to the field components that propagates away from the IDT at either end. For an arbitrary number of coupling points these equations generalizes to,
\begin{align}
\phi_J & = -\frac{ip_J}{\omega C_{\Sigma}} +\frac{C^2_cZ_0p_JH_n(\omega)}{2C_{\Sigma}^2} \label{syseq1}\\
 & +  \frac{C_c}{C_{\Sigma}}\left(\phi^{\text{in}}_0 + \phi^{\text{in}}_d \right)A_n(\omega), \nonumber \\
 p_J & = \frac{i\phi_J}{\omega L_J}, \\
 \phi_0^{\text{out}} &= \phi^{\text{in}}_de^{-i \omega \tau n}  + \frac{Z_0C_c p_J A_n(\omega)}{2C_{\Sigma}} \label{output}, \\
 \phi^{\text{out}}_d& = \phi^{\text{in}}_0e^{-i \omega \tau n } + \frac{Z_0C_c p_JA_n(\omega)}{2C_{\Sigma}},
\label{syseq2}
\end{align}
where $\phi_0$ and $\phi_d$ still represents the two outer coupling points out of $n$ total, $C_{\Sigma} = nC_c + C_J$ is the total capacitance seen by the Josephson junction, and we introduced two frequency-dependent factors
\begin{align}
A_n(\omega) &= \sum_{k=0}^{n-1} e^{-i \omega \tau k}, \\
H_n(\omega) &= n + \sum_{k=1}^{n-1}2ke^{-i \omega \tau (n-k)}.
\label{H}
\end{align}
The first term $A_n(\omega)$ is the so called array factor in classical descriptions of IDTs \cite{Datta,Morgan2007}. In our final set of equations \eqref{syseq1}-\eqref{syseq2}, the part of the transmission line covered by the IDT is effectively traced out and the coupling is reduced to a single coupling point, with in and outgoing fields at the two sides. The extended nature of the interdigitated coupling is accounted for in the frequency dependence of $H_n$ and $A_n$. We note that our formalism allows us to solve for the field components inside the IDT as well, if wanted. The two frequency-dependent factors $H_n$ and $A_n$ are in fact related to each other,
\begin{equation}
\text{Re}[H_n] = |A_n|^2.
\end{equation}
The real and imaginary part of $H_n$ are also related to each other by a Hilbert transform. This connection comes as no surprise as it stems from the Kramers-Kronig relation. Around the center frequency of the IDT, the real and imaginary part of $H_n$ can be well approximated by a Sinc function squared and its Hilbert transform,
\begin{align}
\text{Re}[H_n] &\approx n^2\text{sinc}^2\left(X\right), \label{realH} \\
\text{Im}[H_n] &\approx \frac{n^2(\sin(2X) - 2X)}{2X^2}, \label{imagH} \\
X &= \frac{n \pi(\omega - \omega_{IDT})}{\omega_{IDT}}. 
\end{align}
As will be seen explicitly later on, Eq.~\eqref{realH} enters into the decay rate of the atom, and Eq.~\eqref{imagH} is responsible for either shifting the resonance frequency of the atom, which happens when the atom is not on resonance with the IDT, $\omega_0 \neq \omega_{\text{IDT}}$, or gives rise to vacuum-Rabi splitting, which happens when the atom is strongly coupled and on resonance with the IDT, $\omega_0 = \omega_{\text{IDT}}$. We plot $\text{Re}[H_n]$ and $\text{Im}[H_n]$ for $n=10$, corresponding to an IDT with 10 single finger pairs, in Fig.~\ref{GandB}. 

\section{Results}
\label{results}
In this section investigate the response of the SAW-coupled atom in various coupling regimes. We specifically study the case of an IDT with 10 finger pairs on top of GaAs and LiNbO$_3$ respectively, and derive a $\textit{cavity criterion}$ for our system. 
\subsection{Charge response}
The system dynamics can be quantified in terms of the linear response of the charge operator $p_J$, corresponding to the charge on the Josephson junction. The charge response is obtained as a function of an incoming field, $\phi^{\text{in}} = \overrightarrow{\phi}(0^-)$, from the acoustic transmission line by solving for $p_J$ in equation \eqref{syseq1},
\begin{equation}
\begin{split}
p_J &= \frac{C_cA_n\omega \phi^{\text{in}}}{LC_{\Sigma}(\omega^2 - \omega_0^2 - i\gamma_n(\omega) \omega)} \\
 & \equiv \chi_n(\omega)\frac{ V^{\text{in}}}{L_J}
 \label{pj}
\end{split}
\end{equation}
where $\omega_0^2 = 1/LC_{\Sigma}$ is the bare atom frequency, and $\gamma_n(\omega) = Z_n(\omega)/L_J = \frac{Z_0C_c^2}{2L C_{\Sigma}^2}H_n(\omega)$ is a complex and frequency dependent damping factor. We will refer to the damping factor as a decay rate, since it corresponds to the decay rate of a single excitation from the $\ket{1} \rightarrow \ket{0}$ state in the non-linear, atomic, case. The response function $\chi_n$ describes precisely the charge response of a series  RLC-circuit, where the oscillator is damped by an effective complex and frequency-dependent impedance $Z_n(\omega)$, driven by an incoming voltage field $V^{\text{in}} = C_cA_n\omega \phi^{\text{in}}/C_{\Sigma}$. 

\begin{figure}
\includegraphics[]{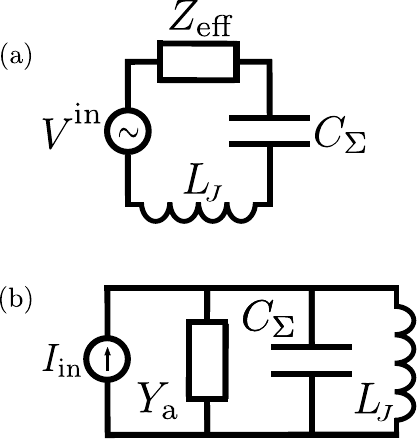}
\caption{\label{effective_circuit} (a) Effective series circuit representation of a SAW-coupled atom in the linear, single excitation, regime. (b) Parallel representation of the circuit in a). The current at the Josephson node is now given by the current through the inductor branch.  }
\end{figure}

The expression for $p_J$ in Eq.~\eqref{pj} is exactly mappable to a parallel RLC-circuit, often used to simulate the behavior of an IDT, and which has been used to interpret several experiments \cite{Gustafsson2014,Moores2017}. The connection becomes clear if we take the time derivative of Eq.~\eqref{pj} and rewrite it in terms of admittances,
\begin{equation}
i\omega p_J  = \frac{Y_L}{Y_L + Y_{C_{\Sigma}} + Y_{\text{a}}} i \omega p_{\text{in}}, 
\end{equation}
where $Y_L = i/\omega L$, $Y_{C_{\Sigma}} = i\omega C_{\Sigma}$, $Y_{\text{a}} = Z_0 C_c^2\omega_0^2H_n(\omega)/2$, and $p_{\text{in}} = C_cA_n(\omega)\phi^{\text{in}}\omega$. This equation can be interpreted as the current passing through the inductor branch of a parallel RLC-circuit, driven by a current source $i\omega p_{\text{in}}$. 
The two equivalent circuits derived from our model is depicted in Fig.~\ref{effective_circuit}. 


\subsection{Acoustic and electric reflection coefficient}
From the input-output relations in Eq.~\eqref{output} we obtain an acoustic reflection coefficient $r^{\text{ac}}_n = \phi_0^{\text{out}}/\phi_0^{\text{in}}$,
\begin{equation}
r^{\text{ac}}_n =i\frac{\text{Re}[\gamma_n(\omega)]\omega}{\omega^2 - \omega_0^2 - i\gamma_n(\omega) \omega}.
\label{ref_final}
\end{equation}
Because all acoustic signals used to probe the system are filtered through an IDT, usually giving a very limited probing bandwidth, it can be advantageous to also probe the system via an electric gate. We incorporate the effect of an electric gate in our model by coupling a second transmission line to the atom, through a small capacitance $C_g$. The gate is coupled directly to the Josephson junction, and the derivation in section \ref{model} is repeated. The gate reflection coefficient can then be obtained in terms of an incoming microwave field from the gate, $r^{\text{g}}_n = \phi^{\text{out}}_g/\phi^{\text{in}}_g$,
\begin{equation}
r^{\text{g}}_n = 1 + i\frac{2\gamma_g \omega }{\omega^2-\omega_0^2 - i(\gamma_n(\omega) + \gamma_g)\omega},
\end{equation} 
where $\gamma_g = \frac{C_g^2Z_{\text{el}}}{LC_{\Sigma}^2}$, $Z_{\text{el}}$ denotes the impedance of the microwave transmission line, which is usually around $50$ $\Omega$, and $C_{\Sigma} = nC_c + C_J + C_g$. Another coefficient of experimental interest that can easily be obtained from our calculations is the transduction coefficient $t^{\text{ac/g}} = \phi_0^{\text{out}}/\phi_g^{\text{in}}$, i.e. what is emitted acoustically when the qubit is driven coherently from the gate,
\begin{equation}
t^{\text{ac/g}} = \frac{i2C_cC_gZ_0A_n(\omega)\omega}{LC_{\Sigma}^2( \omega^2-\omega_0^2 - i(\gamma_n(\omega) + \gamma_g) \omega}.
\end{equation}

\subsection{IDT as a cavity}
The denominator of the charge response $\chi_n$ in Eq.~\eqref{pj} determines the resonance behavior of the system. The same factor, $\omega^2-\omega_0^2 - i\gamma_n(\omega)\omega$, shows up in the denominator of both the acoustic and the electric reflection coefficients. For a fixed $\omega_0$, it is the amplitude of $\gamma_n(\omega_0)$ that determines whether the system exhibits a single resonance, or two resonances due to vacuum-Rabi splitting. By extracting the $n^2$ dependence in the amplitude of $H_n$, the amplitude of the decay rate becomes,
\begin{equation}
\gamma_0 = \frac{Z_0C_c^2\omega_0^2n^2}{2C_{\Sigma}}.
\label{damping}
\end{equation}
However, the capacitance of the Josephson junction is small compared to the IDT capacitance, we can thus make the approximation $C_{\Sigma} = nC_c + C_J \approx nC_c $. Using this approximation, and then inserting the SAW-parameters from section \ref{ingredients} into $Z_0$ and $C_c$, we arrive at the normalized decay rate,
\begin{equation}
\frac{\gamma_0}{\omega_0} = 0.5n K^2,
\label{gamma}
\end{equation} 
where we also made the approximation $\omega_0^2/\omega \approx \omega_0$. Thus, for a fixed atom frequency, the decay rate scales linearly with respect to the piezoelectric coupling constant, and the number of finger pairs in the IDT. 

To enter the regime where the IDT acts as a cavity for the atom, the decay rate needs to be large enough such that an excitation can be interchanged between the atom and the IDT multiple times before it leaves the system at either end. Mathematically, we find two poles in the complex plane of $\chi_n(\omega)$ when the system fulfills the criterion $\gamma_0T_0 \geq 1$, where $T_0 = n\tau$ is the phononic traveling time across the IDT. Using the decay rate from Eq.~\eqref{gamma}, this criterion, as we will call the \textit{cavity criterion}, can be expressed as,
\begin{equation}
0.5\pi K^2 n^2 \geq 1,
\label{criteria}
\end{equation}
where we used that $\tau = 2 \pi/\omega_0$, which is true when the IDT and the atom are on resonance $\omega_{\text{IDT}} = \omega_0$. The behavior of the IDT is thus only determined by the substrate dependent parameter $K^2$, and the number of finger pairs in the IDT $n$.

The factor of $0.5$ in Eq.~\eqref{gamma} and Eq.~\eqref{criteria} should be taken as an approximate value. It is determined by our choice of coupling capacitance, and can be seen as a geometric factor decided by the shape of the IDT. In a lumped element treatment of the IDT, as was done in \cite{Gustafsson2014}, it takes the value $\approx 0.6$, which yields a slightly lower bound on the number of coupling points needed to enter the cavity regime. 

Although $K^2$ cannot be changed \textit{in situ}, it varies dramatically between a weakly piezoelectric material such as GaAs, $K^2 = 0.07\%$, and a strongly piezoelectric material such as LiNbO$_3$, $K^2 = 4.8 \%$ \cite{Morgan2007}. The number of fingers in the IDT is at first sight a more flexible parameter, but it is in fact limited by the $E_J/E_c$ ratio. To many fingers decreases the charging energy of the transmon, and on strongly piezoelectric materials, such as LiNbO$_3$, the anharmonicity can become smaller than the coupling strength, which might be an unfavorable regime for many quantum applications. Thus, to exemplify the two possible system regimes we will study an atom with a fixed number of IDT fingers, and vary the coupling strength by changing the substrate material. We will study the case of 10 coupling points, corresponding to a 20 single finger IDT, on top of GaAs and LiNbO$_3$ respectively. According to the cavity criterion in  Eq.~\eqref{criteria}, we expect the atom on GaAs to exhibit a single resonance frequency, whereas the atom on LiNbO$_3$ is expected to hybridize and exhibit a double resonance behavior. 

\begin{figure*}
\includegraphics[]{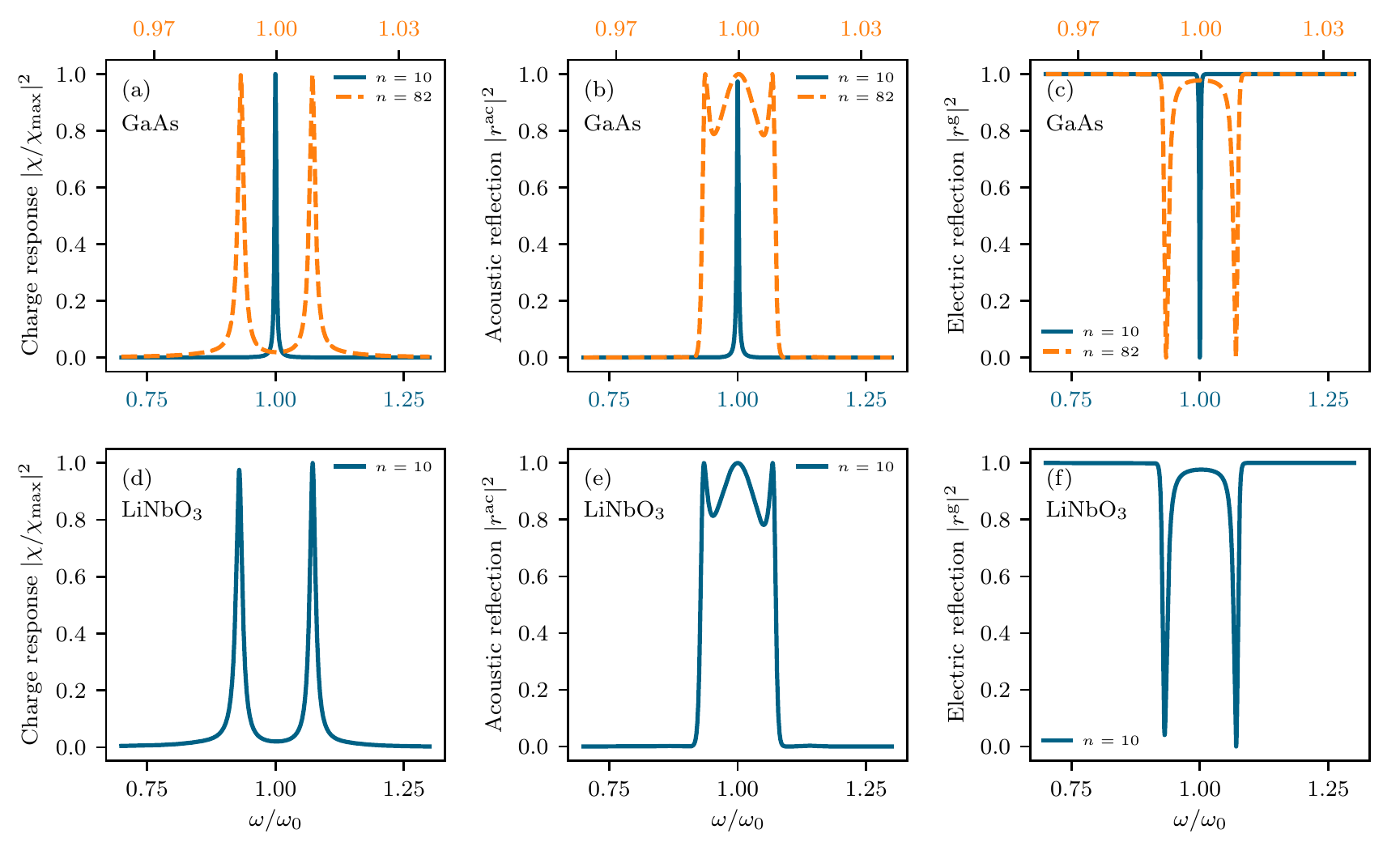}
\caption{\label{ref} Charge response of the Josephson junction and reflection coefficients plotted as a function of frequency for an atom on GaAs in (a)-(c) and LiNbO$_3$ in (d)-(f). The system dynamics exhibits vacuum-Rabi splitting if the cavity criterion in Eq.~\eqref{criteria} is fulfilled. Since the atom on LiNbO$_3$ with $n=10$ (d)-(f) and an atom on GaAs with $n=82$, orange dashed line in (a)-(c), have the same left-hand side of the cavity criterion, their spectra are identical, except that the atom on GaAs is more narrow in frequency since the IDT bandwidth scales inversely with $n$. (a) The charge response of an atom on GaAs with $n=10$ (blue) shows a single resonance peak at the bare atom frequency, as expected for an atom coupled directly to an open transmission line. This resonance is trivially probed both acoustically (b) and electrically (c). On LiNbO$_3$ with $n=10$ the charge response (d) shows two dressed resonances as a result of vacuum-Rabi splitting. The acoustic reflection coefficient (e) shows in addition to the two dressed resonances, a broad peak in the middle of the spectrum. This is not a true atomic resonance of the system but an effect of off-resonant scattering against the two neighboring atomic resonances and direct scattering against the IDT (the two atomic resonances have a small overlap at $\omega_0$ that is hard to see because we plot the magnitude squared). Bypassing the IDT by probing the system electrically (f) removes this peak.} 
\end{figure*}
In Fig.~\ref{ref}(a) and \ref{ref}d we plot the charge response $\chi_{10}$, Eq.~\eqref{pj}, for GaAs and LiNbO$_3$ respectively. On GaAs, the atom has a single Lorentzian resonance peak (blue line), located at the bare atom frequency. The peak has a well defined full width at half maximum (FWHM), given purely by the real part of $\gamma_n(\omega)$, evaluated at $\omega_0$, $\text{FWHM}/\omega_0 = \text{Re}[\gamma_{10}(\omega_0)]/\omega_0 \approx 0.4 \% $. The imaginary part of $\gamma_n(\omega)$ does not alter the dynamics in this case due to the small coupling strength. Contrarily, on LiNbO$_3$, the charge response shows the expected double peak structure. The line shape is no longer Lorentzian and the FWHM is not given purely by the real part of $\gamma_{10}(\omega_0)$. The maximum damping of the atom put the system in the USC regime, $\gamma_{10}(\omega_0)/\omega_0 \approx 23\%$. However, we note that the system is different from an atom ultra-strongly coupled to a microwave transmission line. In our case the system dynamics have changed from an atom coupled directly to a transmission line, to an atom coupled to a cavity, which in turn is coupled to a transmission line. The two resulting dressed resonances are not ultra strongly coupled to the environment, and have a FWHM less than $10 \%$, which is predicted to be the limit at which the effects of the USC appears \cite{Sanchez-Burillo2014,Forn-Diaz2016,Diaz-Camacho2016}. To show that the cavity regime can be entered by increasing the number of fingers as well, we plot the response on GaAs with $n=82$ in Fig.~\ref{ref}(a) (orange dashed line), and the corresponding acoustic and electric reflection coefficients in Fig.~\ref{ref}(b) and Fig.~\ref{ref}(c) respectively. Since the left-hand side of the cavity criterion is identical for $n=82$ on GaAs and $n=10$ on LiNbO$_3$, the resulting dynamics is identical for the two cases. The only difference is a narrowing in frequency for $n=82$ since the IDT bandwidth scales inversely with $n$.

Looking at the real part of the denominator of the charge response in Eq.~\eqref{pj}, it can be surprising that the cavity regime only exhibits two resonances, and not three. The equation $\omega^2-\omega_0^2  + \text{Im}[\gamma_n(\omega)]\omega = 0$ does indeed have three solutions, as long as the cavity criterion is fulfilled. However, one of the solutions is at the bare atom frequency $\omega_0$, but that solutions is suppressed by a maximum in $\text{Re}[\gamma_n(\omega)]$ at $\omega_0$. Interestingly, the exact same mathematical condition occurs for a two level system coupled to a continuum with a modified density of states due the presence of a Fabry-P\'erot cavity, as was pointed out in \cite{Krimer2014}, which further demonstrates the similarities between an IDT and a cavity in this regime. 

In Fig.~\ref{ref}(b) and \ref{ref}(e) we plot the acoustic reflection coefficients, Eq.~\eqref{ref_final}, for GaAs and LiNbO$_3$ respectively, corresponding to the coherent spectrum of a weakly driven atom. On GaAs, acoustic spectroscopy shows precisely the resonance peak at $\omega_0$ as we expect from the charge response in Fig.~\ref{ref}(a). On LiNbO$_3$ however, the acoustic reflection shows, in addition to the two resonance peaks of Fig.~\ref{ref}(d), a broad feature located at $\omega_0$. This resonance should not be mistaken for an atomic resonance of the system. It is partly due to off-resonant scattering against the two neighboring atomic resonances, and direct scattering against the IDT. This feature is unavoidable for acoustic atoms as long as the cavity criterion is fulfilled. In the limit of infinite coupling strength, $\{K^2,n\} \rightarrow \infty$, the two atomic resonances do not overlap anymore and their contribution to the reflected field vanishes, but simultaneously, the IDT alone will fully reflect the incoming signal at $\omega_0$, in the spirit of a Bragg reflector. Thus, we can make the general statement: an atom on resonance with its coupling IDT will fully reflect an incoming coherent field at $\omega_0$ in all parameter regimes. The field can either be reflected entirely from the atom,  entirely from the IDT, or as a combination thereof.

Finally, we plot the electric reflection coefficient in Fig.~\ref{ref}(c) and \ref{ref}(f) for GaAs and LiNbO$_3$ respectively. On GaAs, the electric reflection looks identical to the acoustic reflection, but inverted, since all signals are now reflected unless they are converted into SAW. On LiNbO$_3$ however, the electric reflection allows us to bypass the IDT, and as expected, the direct scattering against the IDT, which resulted in the broad peak at $\omega_0$ in Fig.~\ref{ref}(e), is gone. This confirms our earlier claim that there are only two true atomic resonances.

\subsection{Anti-crossing between the atom and the IDT}
\begin{figure}
\includegraphics{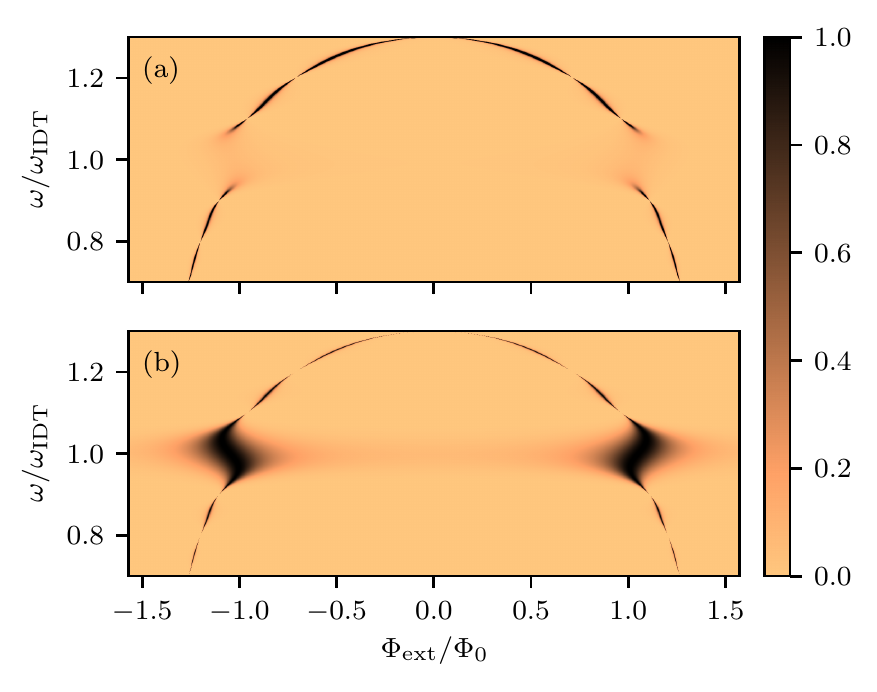}
\caption{\label{fluxtune} Magnitude squared of the electric gate reflection coefficient (a) and the acoustic reflection coefficient (b) on LiNbO$_3$ for $n=10$, as the atom is tuned in and out of resonance with IDT by an external flux. Note that the color coding is inverted and normalized in the electric case for easier comparison with the acoustic spectrum (the gate reflects the incoming field unless its on resonance with the atom, in which case a small part of the energy is converted into SAWs). An avoided crossing can be seen between the atom and the IDT in the electric gate reflection. In the acoustic reflection, the gap is closed due to a combination of direct scattering against the IDT and scattering against the two neighboring atomic resonances.}
\end{figure}

The atom frequency can be tuned by turning the Josephson junction into a SQUID. In the linear regime that we are considering this implies a flux-tunable inductance,
\begin{equation}
L_J(\Phi_{\text{ext}}) = \frac{\Phi_0^2}{4 \pi^2 E_J \cos(\frac{2 \pi \Phi_{\text{ext}}}{\Phi_0})},
\end{equation}
where $\Phi_{\text{ext}}$ is the external field penetrating the SQUID loop. According to the analysis of the previous section, we would expect to see an avoided crossing between the atom and IDT in the \textit{cavity regime}, as the atom is tuned in and out of resonance with the IDT center-frequency. This is precisely what we see in Fig.~\ref{fluxtune}(a), where we plot the magnitude squared of the electric reflection coefficient $r_g^{\mathrm{el}}$, as a function of an external flux. When the atom is on resonance with the IDT, which happens close to $\Phi_{\text{ext}}/\Phi_0 = \pm 1.1$, an avoided crossing occurs. In the acoustic reflection, plotted in Fig.~\ref{fluxtune}(b), the avoided crossing cannot be seen, as the atom-assisted IDT reflection closes the gap once the atom and IDT become resonant. Figure \ref{fluxtune} shows that the electric gate reflection offers a more direct probe of the system dynamics.  

\section{conclusion}

We have shown that the coupling IDT of a SAW-coupled artificial atom can act as a cavity. The situation occurs for artificial atoms on top of strongly piezoelectric materials, such as LiNbO$_3$, or for atoms on any material with sufficient number of IDT-fingers. As long as the \textit{cavity criterion}, $0.5\pi K^2 n^2 \geq 1$, is fulfilled, the system dynamics manifest the characteristic vacuum-Rabi splitting associated with the hybridization of an atom and a cavity. Since the IDT constitutes a major part of the atom itself, the atom can be regarded as forming its own cavity. This exotic behavior results from the slow propagation speed of SAWs, together with the spatial extension of the interdigitated coupling. The dynamics is very similar to other quantum systems with large time delays, such as the giant atom \cite{Guo2017}. 

To verify our claims, we have calculated the system's response to a weak coherent probe, both acoustic and electric, and have shown that an avoided crossing between the atom and IDT can be seen as the atom is tuned in and out of resonance with the IDT center-frequency. Based on our calculations, we propose electric gate reflection measurements as the best experimental way to probe the system dynamics, as an acoustic reflection measurement would show additional features due to direct scattering against the IDT alone. Moreover, acoustic reflection measurements can typically only cover a narrow frequency band.

Additionally, we have mapped important SAW-parameters onto circuit elements, enabling standard circuit quantization techniques from cQED to be carried over to the field of cQAD. Our model offers a bridge between earlier phenomenological models of the acoustic atom \cite{FriskKockum2014,Guo2017}, and the purely classical lumped circuit models used to interpret several experiments \cite{Gustafsson2014,Moores2017}.

We believe our findings are of special importance regarding cQAD and the ultrastrong coupling regime. Our calculations show that an artificial atom on LiNbO$_3$ cannot be regarded as an atom ultra-strongly coupled to a transmission line, since the coupling-IDT acts as a cavity in this regime. We have also demonstrated that in order to address non-linear quantum effects of an atom coupled to SAW in this regime, one would need theoretical methods that can deal with both ultrastrong coupling and time delays. We do not know of any methods that combine these two effects in a satisfactory way. Thus, developing such a formalism would be an interesting pursuit in future studies. 
\label{conclusion}

\begin{acknowledgments}
We would like to thank Thomas Aref, Anton Frisk Kockum, and Vitaly Shumeiko for fruitful discussions. We also acknowledge financial support from the Swedish Research Council and the Knut and Alice Wallenberg Foundation.
\end{acknowledgments}

\bibliography{cavityfree.bib}

\begin{thebibliography}{36}%
\makeatletter
\providecommand \@ifxundefined [1]{%
 \@ifx{#1\undefined}
}%
\providecommand \@ifnum [1]{%
 \ifnum #1\expandafter \@firstoftwo
 \else \expandafter \@secondoftwo
 \fi
}%
\providecommand \@ifx [1]{%
 \ifx #1\expandafter \@firstoftwo
 \else \expandafter \@secondoftwo
 \fi
}%
\providecommand \natexlab [1]{#1}%
\providecommand \enquote  [1]{``#1''}%
\providecommand \bibnamefont  [1]{#1}%
\providecommand \bibfnamefont [1]{#1}%
\providecommand \citenamefont [1]{#1}%
\providecommand \href@noop [0]{\@secondoftwo}%
\providecommand \href [0]{\begingroup \@sanitize@url \@href}%
\providecommand \@href[1]{\@@startlink{#1}\@@href}%
\providecommand \@@href[1]{\endgroup#1\@@endlink}%
\providecommand \@sanitize@url [0]{\catcode `\\12\catcode `\$12\catcode
  `\&12\catcode `\#12\catcode `\^12\catcode `\_12\catcode `\%12\relax}%
\providecommand \@@startlink[1]{}%
\providecommand \@@endlink[0]{}%
\providecommand \url  [0]{\begingroup\@sanitize@url \@url }%
\providecommand \@url [1]{\endgroup\@href {#1}{\urlprefix }}%
\providecommand \urlprefix  [0]{URL }%
\providecommand \Eprint [0]{\href }%
\providecommand \doibase [0]{http://dx.doi.org/}%
\providecommand \selectlanguage [0]{\@gobble}%
\providecommand \bibinfo  [0]{\@secondoftwo}%
\providecommand \bibfield  [0]{\@secondoftwo}%
\providecommand \translation [1]{[#1]}%
\providecommand \BibitemOpen [0]{}%
\providecommand \bibitemStop [0]{}%
\providecommand \bibitemNoStop [0]{.\EOS\space}%
\providecommand \EOS [0]{\spacefactor3000\relax}%
\providecommand \BibitemShut  [1]{\csname bibitem#1\endcsname}%
\let\auto@bib@innerbib\@empty
\bibitem [{\citenamefont {Blais}\ \emph {et~al.}(2004)\citenamefont {Blais},
  \citenamefont {Huang}, \citenamefont {Wallraff}, \citenamefont {Girvin},\
  and\ \citenamefont {Schoelkopf}}]{Blais2004}%
  \BibitemOpen
  \bibfield  {author} {\bibinfo {author} {\bibfnamefont {Alexandre}\
  \bibnamefont {Blais}}, \bibinfo {author} {\bibfnamefont {Ren-Shou}\
  \bibnamefont {Huang}}, \bibinfo {author} {\bibfnamefont {Andreas}\
  \bibnamefont {Wallraff}}, \bibinfo {author} {\bibfnamefont {S.~M.}\
  \bibnamefont {Girvin}}, \ and\ \bibinfo {author} {\bibfnamefont {R.~J.}\
  \bibnamefont {Schoelkopf}},\ }\bibfield  {title} {\enquote {\bibinfo {title}
  {{Cavity quantum electrodynamics for superconducting electrical circuits: An
  architecture for quantum computation}},}\ }\href {\doibase
  10.1103/PhysRevA.69.062320} {\bibfield  {journal} {\bibinfo  {journal}
  {Physical Review A}\ }\textbf {\bibinfo {volume} {69}},\ \bibinfo {pages}
  {062320} (\bibinfo {year} {2004})}\BibitemShut {NoStop}%
\bibitem [{\citenamefont {Wallraff}\ \emph {et~al.}(2004)\citenamefont
  {Wallraff}, \citenamefont {Schuster}, \citenamefont {Blais}, \citenamefont
  {Frunzio}, \citenamefont {Huang}, \citenamefont {Majer}, \citenamefont
  {Kumar}, \citenamefont {Girvin},\ and\ \citenamefont
  {Schoelkopf}}]{Wallraff2004}%
  \BibitemOpen
  \bibfield  {author} {\bibinfo {author} {\bibfnamefont {A.}~\bibnamefont
  {Wallraff}}, \bibinfo {author} {\bibfnamefont {D.~I.}\ \bibnamefont
  {Schuster}}, \bibinfo {author} {\bibfnamefont {A.}~\bibnamefont {Blais}},
  \bibinfo {author} {\bibfnamefont {L.}~\bibnamefont {Frunzio}}, \bibinfo
  {author} {\bibfnamefont {R.-S.}\ \bibnamefont {Huang}}, \bibinfo {author}
  {\bibfnamefont {J.}~\bibnamefont {Majer}}, \bibinfo {author} {\bibfnamefont
  {S.}~\bibnamefont {Kumar}}, \bibinfo {author} {\bibfnamefont {S.~M.}\
  \bibnamefont {Girvin}}, \ and\ \bibinfo {author} {\bibfnamefont {R.~J.}\
  \bibnamefont {Schoelkopf}},\ }\bibfield  {title} {\enquote {\bibinfo {title}
  {{Strong coupling of a single photon to a superconducting qubit using circuit
  quantum electrodynamics}},}\ }\href {\doibase 10.1038/nature02851} {\bibfield
   {journal} {\bibinfo  {journal} {Nature}\ }\textbf {\bibinfo {volume}
  {431}},\ \bibinfo {pages} {162--167} (\bibinfo {year} {2004})}\BibitemShut
  {NoStop}%
\bibitem [{\citenamefont {Gu}\ \emph {et~al.}(2017)\citenamefont {Gu},
  \citenamefont {Kockum}, \citenamefont {Miranowicz}, \citenamefont {Liu},\
  and\ \citenamefont {Nori}}]{Gu2017}%
  \BibitemOpen
  \bibfield  {author} {\bibinfo {author} {\bibfnamefont {Xiu}\ \bibnamefont
  {Gu}}, \bibinfo {author} {\bibfnamefont {Anton~Frisk}\ \bibnamefont
  {Kockum}}, \bibinfo {author} {\bibfnamefont {Adam}\ \bibnamefont
  {Miranowicz}}, \bibinfo {author} {\bibfnamefont {Yu-xi}\ \bibnamefont {Liu}},
  \ and\ \bibinfo {author} {\bibfnamefont {Franco}\ \bibnamefont {Nori}},\
  }\bibfield  {title} {\enquote {\bibinfo {title} {{Microwave photonics with
  superconducting quantum circuits}},}\ }\href {\doibase
  10.1016/j.physrep.2017.10.002} {\bibfield  {journal} {\bibinfo  {journal}
  {Physics Reports}\ }\textbf {\bibinfo {volume} {718}},\ \bibinfo {pages}
  {1--102} (\bibinfo {year} {2017})}\BibitemShut {NoStop}%
\bibitem [{\citenamefont {Wendin}(2017)}]{Wendin2017}%
  \BibitemOpen
  \bibfield  {author} {\bibinfo {author} {\bibfnamefont {G}~\bibnamefont
  {Wendin}},\ }\bibfield  {title} {\enquote {\bibinfo {title} {{Quantum
  information processing with superconducting circuits: a review}},}\ }\href
  {\doibase 10.1088/1361-6633/aa7e1a} {\bibfield  {journal} {\bibinfo
  {journal} {Reports on Progress in Physics}\ }\textbf {\bibinfo {volume}
  {80}},\ \bibinfo {pages} {106001} (\bibinfo {year} {2017})}\BibitemShut
  {NoStop}%
\bibitem [{\citenamefont {Gustafsson}\ \emph {et~al.}(2014)\citenamefont
  {Gustafsson}, \citenamefont {Aref}, \citenamefont {Kockum}, \citenamefont
  {Ekstr{\"{o}}m}, \citenamefont {Johansson},\ and\ \citenamefont
  {Delsing}}]{Gustafsson2014}%
  \BibitemOpen
  \bibfield  {author} {\bibinfo {author} {\bibfnamefont {Martin~V}\
  \bibnamefont {Gustafsson}}, \bibinfo {author} {\bibfnamefont {Thomas}\
  \bibnamefont {Aref}}, \bibinfo {author} {\bibfnamefont {Anton~Frisk}\
  \bibnamefont {Kockum}}, \bibinfo {author} {\bibfnamefont {Maria~K}\
  \bibnamefont {Ekstr{\"{o}}m}}, \bibinfo {author} {\bibfnamefont
  {G{\"{o}}ran}\ \bibnamefont {Johansson}}, \ and\ \bibinfo {author}
  {\bibfnamefont {Per}\ \bibnamefont {Delsing}},\ }\bibfield  {title} {\enquote
  {\bibinfo {title} {{Propagating phonons coupled to an artificial atom.}}}\
  }\href {\doibase 10.1126/science.1257219} {\bibfield  {journal} {\bibinfo
  {journal} {Science (New York, N.Y.)}\ }\textbf {\bibinfo {volume} {346}},\
  \bibinfo {pages} {207--11} (\bibinfo {year} {2014})}\BibitemShut {NoStop}%
\bibitem [{\citenamefont {Manenti}\ \emph {et~al.}(2017)\citenamefont
  {Manenti}, \citenamefont {Kockum}, \citenamefont {Patterson}, \citenamefont
  {Behrle}, \citenamefont {Rahamim}, \citenamefont {Tancredi}, \citenamefont
  {Nori},\ and\ \citenamefont {Leek}}]{Manenti2017}%
  \BibitemOpen
  \bibfield  {author} {\bibinfo {author} {\bibfnamefont {Riccardo}\
  \bibnamefont {Manenti}}, \bibinfo {author} {\bibfnamefont {Anton~F.}\
  \bibnamefont {Kockum}}, \bibinfo {author} {\bibfnamefont {Andrew}\
  \bibnamefont {Patterson}}, \bibinfo {author} {\bibfnamefont {Tanja}\
  \bibnamefont {Behrle}}, \bibinfo {author} {\bibfnamefont {Joseph}\
  \bibnamefont {Rahamim}}, \bibinfo {author} {\bibfnamefont {Giovanna}\
  \bibnamefont {Tancredi}}, \bibinfo {author} {\bibfnamefont {Franco}\
  \bibnamefont {Nori}}, \ and\ \bibinfo {author} {\bibfnamefont {Peter~J.}\
  \bibnamefont {Leek}},\ }\bibfield  {title} {\enquote {\bibinfo {title}
  {{Circuit quantum acoustodynamics with surface acoustic waves}},}\ }\href
  {\doibase 10.1038/s41467-017-01063-9} {\bibfield  {journal} {\bibinfo
  {journal} {Nature Communications}\ }\textbf {\bibinfo {volume} {8}},\
  \bibinfo {pages} {975} (\bibinfo {year} {2017})}\BibitemShut {NoStop}%
\bibitem [{\citenamefont {Moores}\ \emph {et~al.}(2017)\citenamefont {Moores},
  \citenamefont {Sletten}, \citenamefont {Viennot},\ and\ \citenamefont
  {Lehnert}}]{Moores2017}%
  \BibitemOpen
  \bibfield  {author} {\bibinfo {author} {\bibfnamefont {Bradley~A.}\
  \bibnamefont {Moores}}, \bibinfo {author} {\bibfnamefont {Lucas~R.}\
  \bibnamefont {Sletten}}, \bibinfo {author} {\bibfnamefont {Jeremie~J.}\
  \bibnamefont {Viennot}}, \ and\ \bibinfo {author} {\bibfnamefont {K.~W.}\
  \bibnamefont {Lehnert}},\ }\bibfield  {title} {\enquote {\bibinfo {title}
  {{Cavity quantum acoustic device in the multimode strong coupling regime}},}\
  }\href {http://arxiv.org/abs/1711.05913} {\  (\bibinfo {year} {2017})},\
  \Eprint {http://arxiv.org/abs/1711.05913} {arXiv:1711.05913} \BibitemShut
  {NoStop}%
\bibitem [{\citenamefont {Noguchi}\ \emph {et~al.}(2017)\citenamefont
  {Noguchi}, \citenamefont {Yamazaki}, \citenamefont {Tabuchi},\ and\
  \citenamefont {Nakamura}}]{Noguchi2017}%
  \BibitemOpen
  \bibfield  {author} {\bibinfo {author} {\bibfnamefont {Atsushi}\ \bibnamefont
  {Noguchi}}, \bibinfo {author} {\bibfnamefont {Rekishu}\ \bibnamefont
  {Yamazaki}}, \bibinfo {author} {\bibfnamefont {Yutaka}\ \bibnamefont
  {Tabuchi}}, \ and\ \bibinfo {author} {\bibfnamefont {Yasunobu}\ \bibnamefont
  {Nakamura}},\ }\bibfield  {title} {\enquote {\bibinfo {title}
  {{Qubit-Assisted Transduction for a Detection of Surface Acoustic Waves near
  the Quantum Limit}},}\ }\href {\doibase 10.1103/PhysRevLett.119.180505}
  {\bibfield  {journal} {\bibinfo  {journal} {Physical Review Letters}\
  }\textbf {\bibinfo {volume} {119}},\ \bibinfo {pages} {180505} (\bibinfo
  {year} {2017})}\BibitemShut {NoStop}%
\bibitem [{\citenamefont {Bolgar}\ \emph {et~al.}(2018)\citenamefont {Bolgar},
  \citenamefont {Zotova}, \citenamefont {Kirichenko}, \citenamefont {Besedin},
  \citenamefont {Semenov}, \citenamefont {Shaikhaidarov},\ and\ \citenamefont
  {Astafiev}}]{Bolgar2018}%
  \BibitemOpen
  \bibfield  {author} {\bibinfo {author} {\bibfnamefont {Aleksey~N.}\
  \bibnamefont {Bolgar}}, \bibinfo {author} {\bibfnamefont {Julia~I.}\
  \bibnamefont {Zotova}}, \bibinfo {author} {\bibfnamefont {Daniil~D.}\
  \bibnamefont {Kirichenko}}, \bibinfo {author} {\bibfnamefont {Ilia~S.}\
  \bibnamefont {Besedin}}, \bibinfo {author} {\bibfnamefont {Aleksander~V.}\
  \bibnamefont {Semenov}}, \bibinfo {author} {\bibfnamefont {Rais~S.}\
  \bibnamefont {Shaikhaidarov}}, \ and\ \bibinfo {author} {\bibfnamefont
  {Oleg~V.}\ \bibnamefont {Astafiev}},\ }\bibfield  {title} {\enquote {\bibinfo
  {title} {{Quantum Regime of a Two-Dimensional Phonon Cavity}},}\ }\href
  {\doibase 10.1103/PhysRevLett.120.223603} {\bibfield  {journal} {\bibinfo
  {journal} {Physical Review Letters}\ }\textbf {\bibinfo {volume} {120}},\
  \bibinfo {pages} {223603} (\bibinfo {year} {2018})}\BibitemShut {NoStop}%
\bibitem [{\citenamefont {Satzinger}\ \emph {et~al.}(2018)\citenamefont
  {Satzinger}, \citenamefont {Zhong}, \citenamefont {Chang}, \citenamefont
  {Peairs}, \citenamefont {Bienfait}, \citenamefont {Chou}, \citenamefont
  {Cleland}, \citenamefont {Conner}, \citenamefont {Dumur}, \citenamefont
  {Grebel}, \citenamefont {Gutierrez}, \citenamefont {November}, \citenamefont
  {Povey}, \citenamefont {Whiteley}, \citenamefont {Awschalom}, \citenamefont
  {Schuster},\ and\ \citenamefont {Cleland}}]{Satzinger2018}%
  \BibitemOpen
  \bibfield  {author} {\bibinfo {author} {\bibfnamefont {K.~J.}\ \bibnamefont
  {Satzinger}}, \bibinfo {author} {\bibfnamefont {Y.~P.}\ \bibnamefont
  {Zhong}}, \bibinfo {author} {\bibfnamefont {H.~S.}\ \bibnamefont {Chang}},
  \bibinfo {author} {\bibfnamefont {G.~A.}\ \bibnamefont {Peairs}}, \bibinfo
  {author} {\bibfnamefont {A.}~\bibnamefont {Bienfait}}, \bibinfo {author}
  {\bibfnamefont {Ming-Han}\ \bibnamefont {Chou}}, \bibinfo {author}
  {\bibfnamefont {A.~Y.}\ \bibnamefont {Cleland}}, \bibinfo {author}
  {\bibfnamefont {C.~R.}\ \bibnamefont {Conner}}, \bibinfo {author}
  {\bibfnamefont {E.}~\bibnamefont {Dumur}}, \bibinfo {author} {\bibfnamefont
  {J.}~\bibnamefont {Grebel}}, \bibinfo {author} {\bibfnamefont
  {I.}~\bibnamefont {Gutierrez}}, \bibinfo {author} {\bibfnamefont {B.~H.}\
  \bibnamefont {November}}, \bibinfo {author} {\bibfnamefont {R.~G.}\
  \bibnamefont {Povey}}, \bibinfo {author} {\bibfnamefont {S.~J.}\ \bibnamefont
  {Whiteley}}, \bibinfo {author} {\bibfnamefont {D.~D.}\ \bibnamefont
  {Awschalom}}, \bibinfo {author} {\bibfnamefont {D.~I.}\ \bibnamefont
  {Schuster}}, \ and\ \bibinfo {author} {\bibfnamefont {A.~N.}\ \bibnamefont
  {Cleland}},\ }\bibfield  {title} {\enquote {\bibinfo {title} {{Quantum
  control of surface acoustic wave phonons}},}\ }\href
  {http://arxiv.org/abs/1804.07308} {\  (\bibinfo {year} {2018})},\ \Eprint
  {http://arxiv.org/abs/1804.07308} {arXiv:1804.07308} \BibitemShut {NoStop}%
\bibitem [{\citenamefont {Chu}\ \emph {et~al.}(2017)\citenamefont {Chu},
  \citenamefont {Kharel}, \citenamefont {Renninger}, \citenamefont {Burkhart},
  \citenamefont {Frunzio}, \citenamefont {Rakich},\ and\ \citenamefont
  {Schoelkopf}}]{Chu2017}%
  \BibitemOpen
  \bibfield  {author} {\bibinfo {author} {\bibfnamefont {Yiwen}\ \bibnamefont
  {Chu}}, \bibinfo {author} {\bibfnamefont {Prashanta}\ \bibnamefont {Kharel}},
  \bibinfo {author} {\bibfnamefont {William~H}\ \bibnamefont {Renninger}},
  \bibinfo {author} {\bibfnamefont {Luke~D}\ \bibnamefont {Burkhart}}, \bibinfo
  {author} {\bibfnamefont {Luigi}\ \bibnamefont {Frunzio}}, \bibinfo {author}
  {\bibfnamefont {Peter~T}\ \bibnamefont {Rakich}}, \ and\ \bibinfo {author}
  {\bibfnamefont {Robert~J}\ \bibnamefont {Schoelkopf}},\ }\bibfield  {title}
  {\enquote {\bibinfo {title} {{Quantum acoustics with superconducting
  qubits.}}}\ }\href {\doibase 10.1126/science.aao1511} {\bibfield  {journal}
  {\bibinfo  {journal} {Science (New York, N.Y.)}\ }\textbf {\bibinfo {volume}
  {358}},\ \bibinfo {pages} {199--202} (\bibinfo {year} {2017})}\BibitemShut
  {NoStop}%
\bibitem [{\citenamefont {Kervinen}\ \emph {et~al.}(2018)\citenamefont
  {Kervinen}, \citenamefont {Rissanen},\ and\ \citenamefont
  {Sillanp{\"{a}}{\"{a}}}}]{Kervinen2018}%
  \BibitemOpen
  \bibfield  {author} {\bibinfo {author} {\bibfnamefont {Mikael}\ \bibnamefont
  {Kervinen}}, \bibinfo {author} {\bibfnamefont {Ilkka}\ \bibnamefont
  {Rissanen}}, \ and\ \bibinfo {author} {\bibfnamefont {Mika}\ \bibnamefont
  {Sillanp{\"{a}}{\"{a}}}},\ }\bibfield  {title} {\enquote {\bibinfo {title}
  {{Interfacing planar superconducting qubits with high overtone bulk acoustic
  phonons}},}\ }\href {\doibase 10.1103/PhysRevB.97.205443} {\bibfield
  {journal} {\bibinfo  {journal} {Physical Review B}\ }\textbf {\bibinfo
  {volume} {97}},\ \bibinfo {pages} {205443} (\bibinfo {year}
  {2018})}\BibitemShut {NoStop}%
\bibitem [{\citenamefont {Schuetz}\ \emph {et~al.}(2015)\citenamefont
  {Schuetz}, \citenamefont {Kessler}, \citenamefont {Giedke}, \citenamefont
  {Vandersypen}, \citenamefont {Lukin},\ and\ \citenamefont
  {Cirac}}]{Schuetz2015}%
  \BibitemOpen
  \bibfield  {author} {\bibinfo {author} {\bibfnamefont {M.~J.~A.}\
  \bibnamefont {Schuetz}}, \bibinfo {author} {\bibfnamefont {E.~M.}\
  \bibnamefont {Kessler}}, \bibinfo {author} {\bibfnamefont {G.}~\bibnamefont
  {Giedke}}, \bibinfo {author} {\bibfnamefont {L.~M.~K.}\ \bibnamefont
  {Vandersypen}}, \bibinfo {author} {\bibfnamefont {M.~D.}\ \bibnamefont
  {Lukin}}, \ and\ \bibinfo {author} {\bibfnamefont {J.~I.}\ \bibnamefont
  {Cirac}},\ }\bibfield  {title} {\enquote {\bibinfo {title} {{Universal
  Quantum Transducers Based on Surface Acoustic Waves}},}\ }\href {\doibase
  10.1103/PhysRevX.5.031031} {\bibfield  {journal} {\bibinfo  {journal}
  {Physical Review X}\ }\textbf {\bibinfo {volume} {5}},\ \bibinfo {pages}
  {031031} (\bibinfo {year} {2015})}\BibitemShut {NoStop}%
\bibitem [{\citenamefont {Shumeiko}(2016)}]{Shumeiko2016}%
  \BibitemOpen
  \bibfield  {author} {\bibinfo {author} {\bibfnamefont {Vitaly~S.}\
  \bibnamefont {Shumeiko}},\ }\bibfield  {title} {\enquote {\bibinfo {title}
  {{Quantum acousto-optic transducer for superconducting qubits}},}\ }\href
  {\doibase 10.1103/PhysRevA.93.023838} {\bibfield  {journal} {\bibinfo
  {journal} {Physical Review A}\ }\textbf {\bibinfo {volume} {93}},\ \bibinfo
  {pages} {023838} (\bibinfo {year} {2016})}\BibitemShut {NoStop}%
\bibitem [{\citenamefont {Pichler}\ \emph {et~al.}(2017)\citenamefont
  {Pichler}, \citenamefont {Choi}, \citenamefont {Zoller},\ and\ \citenamefont
  {Lukin}}]{Pichler2017a}%
  \BibitemOpen
  \bibfield  {author} {\bibinfo {author} {\bibfnamefont {Hannes}\ \bibnamefont
  {Pichler}}, \bibinfo {author} {\bibfnamefont {Soonwon}\ \bibnamefont {Choi}},
  \bibinfo {author} {\bibfnamefont {Peter}\ \bibnamefont {Zoller}}, \ and\
  \bibinfo {author} {\bibfnamefont {Mikhail~D}\ \bibnamefont {Lukin}},\
  }\bibfield  {title} {\enquote {\bibinfo {title} {{Universal photonic quantum
  computation via time-delayed feedback.}}}\ }\href {\doibase
  10.1073/pnas.1711003114} {\bibfield  {journal} {\bibinfo  {journal}
  {Proceedings of the National Academy of Sciences of the United States of
  America}\ }\textbf {\bibinfo {volume} {114}},\ \bibinfo {pages}
  {11362--11367} (\bibinfo {year} {2017})}\BibitemShut {NoStop}%
\bibitem [{\citenamefont {Datta}(1986)}]{Datta}%
  \BibitemOpen
  \bibfield  {author} {\bibinfo {author} {\bibfnamefont {S.}~\bibnamefont
  {Datta}},\ }\href@noop {} {\emph {\bibinfo {title} {Practise-Hall}}}\
  (\bibinfo  {publisher} {Prentice-Hall},\ \bibinfo {year} {1986})\ p.\
  \bibinfo {pages} {208}\BibitemShut {NoStop}%
\bibitem [{\citenamefont {Koch}\ \emph {et~al.}(2007)\citenamefont {Koch},
  \citenamefont {Yu}, \citenamefont {Gambetta}, \citenamefont {Houck},
  \citenamefont {Schuster}, \citenamefont {Majer}, \citenamefont {Blais},
  \citenamefont {Devoret}, \citenamefont {Girvin},\ and\ \citenamefont
  {Schoelkopf}}]{Koch2007}%
  \BibitemOpen
  \bibfield  {author} {\bibinfo {author} {\bibfnamefont {Jens}\ \bibnamefont
  {Koch}}, \bibinfo {author} {\bibfnamefont {Terri~M.}\ \bibnamefont {Yu}},
  \bibinfo {author} {\bibfnamefont {Jay}\ \bibnamefont {Gambetta}}, \bibinfo
  {author} {\bibfnamefont {A.~A.}\ \bibnamefont {Houck}}, \bibinfo {author}
  {\bibfnamefont {D.~I.}\ \bibnamefont {Schuster}}, \bibinfo {author}
  {\bibfnamefont {J.}~\bibnamefont {Majer}}, \bibinfo {author} {\bibfnamefont
  {Alexandre}\ \bibnamefont {Blais}}, \bibinfo {author} {\bibfnamefont {M.~H.}\
  \bibnamefont {Devoret}}, \bibinfo {author} {\bibfnamefont {S.~M.}\
  \bibnamefont {Girvin}}, \ and\ \bibinfo {author} {\bibfnamefont {R.~J.}\
  \bibnamefont {Schoelkopf}},\ }\bibfield  {title} {\enquote {\bibinfo {title}
  {{Charge-insensitive qubit design derived from the Cooper pair box}},}\
  }\href {\doibase 10.1103/PhysRevA.76.042319} {\bibfield  {journal} {\bibinfo
  {journal} {Physical Review A}\ }\textbf {\bibinfo {volume} {76}},\ \bibinfo
  {pages} {042319} (\bibinfo {year} {2007})}\BibitemShut {NoStop}%
\bibitem [{\citenamefont {{Frisk Kockum}}\ \emph {et~al.}(2014)\citenamefont
  {{Frisk Kockum}}, \citenamefont {Delsing},\ and\ \citenamefont
  {Johansson}}]{FriskKockum2014}%
  \BibitemOpen
  \bibfield  {author} {\bibinfo {author} {\bibfnamefont {Anton}\ \bibnamefont
  {{Frisk Kockum}}}, \bibinfo {author} {\bibfnamefont {Per}\ \bibnamefont
  {Delsing}}, \ and\ \bibinfo {author} {\bibfnamefont {G{\"{o}}ran}\
  \bibnamefont {Johansson}},\ }\bibfield  {title} {\enquote {\bibinfo {title}
  {{Designing frequency-dependent relaxation rates and Lamb shifts for a giant
  artificial atom}},}\ }\href {\doibase 10.1103/PhysRevA.90.013837} {\bibfield
  {journal} {\bibinfo  {journal} {Physical Review A}\ }\textbf {\bibinfo
  {volume} {90}},\ \bibinfo {pages} {013837} (\bibinfo {year}
  {2014})}\BibitemShut {NoStop}%
\bibitem [{\citenamefont {Kockum}\ \emph
  {et~al.}(2018{\natexlab{a}})\citenamefont {Kockum}, \citenamefont
  {Johansson},\ and\ \citenamefont {Nori}}]{Kockum2018a}%
  \BibitemOpen
  \bibfield  {author} {\bibinfo {author} {\bibfnamefont {Anton~Frisk}\
  \bibnamefont {Kockum}}, \bibinfo {author} {\bibfnamefont {G{\"{o}}ran}\
  \bibnamefont {Johansson}}, \ and\ \bibinfo {author} {\bibfnamefont {Franco}\
  \bibnamefont {Nori}},\ }\bibfield  {title} {\enquote {\bibinfo {title}
  {{Decoherence-Free Interaction between Giant Atoms in Waveguide Quantum
  Electrodynamics}},}\ }\href {\doibase 10.1103/PhysRevLett.120.140404}
  {\bibfield  {journal} {\bibinfo  {journal} {Physical Review Letters}\
  }\textbf {\bibinfo {volume} {120}},\ \bibinfo {pages} {140404} (\bibinfo
  {year} {2018}{\natexlab{a}})}\BibitemShut {NoStop}%
\bibitem [{\citenamefont {Ekstr{\"{o}}m}\ \emph {et~al.}(2017)\citenamefont
  {Ekstr{\"{o}}m}, \citenamefont {Aref}, \citenamefont {Runeson}, \citenamefont
  {Bj{\"{o}}rck}, \citenamefont {Bostr{\"{o}}m},\ and\ \citenamefont
  {Delsing}}]{Ekstrom2017}%
  \BibitemOpen
  \bibfield  {author} {\bibinfo {author} {\bibfnamefont {Maria~K.}\
  \bibnamefont {Ekstr{\"{o}}m}}, \bibinfo {author} {\bibfnamefont {Thomas}\
  \bibnamefont {Aref}}, \bibinfo {author} {\bibfnamefont {Johan}\ \bibnamefont
  {Runeson}}, \bibinfo {author} {\bibfnamefont {Johan}\ \bibnamefont
  {Bj{\"{o}}rck}}, \bibinfo {author} {\bibfnamefont {Isac}\ \bibnamefont
  {Bostr{\"{o}}m}}, \ and\ \bibinfo {author} {\bibfnamefont {Per}\ \bibnamefont
  {Delsing}},\ }\bibfield  {title} {\enquote {\bibinfo {title} {{Surface
  acoustic wave unidirectional transducers for quantum applications}},}\ }\href
  {\doibase 10.1063/1.4975803} {\bibfield  {journal} {\bibinfo  {journal}
  {Applied Physics Letters}\ }\textbf {\bibinfo {volume} {110}},\ \bibinfo
  {pages} {073105} (\bibinfo {year} {2017})}\BibitemShut {NoStop}%
\bibitem [{\citenamefont {Yurke}\ and\ \citenamefont
  {Denker}(1984)}]{Yurke1984}%
  \BibitemOpen
  \bibfield  {author} {\bibinfo {author} {\bibfnamefont {Bernard}\ \bibnamefont
  {Yurke}}\ and\ \bibinfo {author} {\bibfnamefont {John~S.}\ \bibnamefont
  {Denker}},\ }\bibfield  {title} {\enquote {\bibinfo {title} {{Quantum network
  theory}},}\ }\href {\doibase 10.1103/PhysRevA.29.1419} {\bibfield  {journal}
  {\bibinfo  {journal} {Physical Review A}\ }\textbf {\bibinfo {volume} {29}},\
  \bibinfo {pages} {1419--1437} (\bibinfo {year} {1984})}\BibitemShut {NoStop}%
\bibitem [{\citenamefont {Devoret}(1995)}]{Devoret1995}%
  \BibitemOpen
  \bibfield  {author} {\bibinfo {author} {\bibfnamefont {Michel~H}\
  \bibnamefont {Devoret}},\ }\bibfield  {title} {\enquote {\bibinfo {title}
  {{Quantum fluctuations in electrical circuits}},}\ }\href@noop {} {\bibfield
  {journal} {\bibinfo  {journal} {Les Houches, Session LXIII}\ }\textbf
  {\bibinfo {volume} {7}} (\bibinfo {year} {1995})}\BibitemShut {NoStop}%
\bibitem [{\citenamefont {Stoneley}(1955)}]{Stoneley1955}%
  \BibitemOpen
  \bibfield  {author} {\bibinfo {author} {\bibfnamefont {R.}~\bibnamefont
  {Stoneley}},\ }\bibfield  {title} {\enquote {\bibinfo {title} {{The
  Propagation of Surface Elastic Waves in a Cubic Crystal}},}\ }\href {\doibase
  10.1098/rspa.1955.0230} {\bibfield  {journal} {\bibinfo  {journal}
  {Proceedings of the Royal Society A: Mathematical, Physical and Engineering
  Sciences}\ }\textbf {\bibinfo {volume} {232}},\ \bibinfo {pages} {447--458}
  (\bibinfo {year} {1955})}\BibitemShut {NoStop}%
\bibitem [{\citenamefont {Simon}(1996)}]{Simon1996}%
  \BibitemOpen
  \bibfield  {author} {\bibinfo {author} {\bibfnamefont {Steven~H.}\
  \bibnamefont {Simon}},\ }\bibfield  {title} {\enquote {\bibinfo {title}
  {{Coupling of surface acoustic waves to a two-dimensional electron gas}},}\
  }\href {\doibase 10.1103/PhysRevB.54.13878} {\bibfield  {journal} {\bibinfo
  {journal} {Physical Review B}\ }\textbf {\bibinfo {volume} {54}},\ \bibinfo
  {pages} {13878--13884} (\bibinfo {year} {1996})}\BibitemShut {NoStop}%
\bibitem [{\citenamefont {Morgan}(2007)}]{Morgan2007}%
  \BibitemOpen
  \bibfield  {author} {\bibinfo {author} {\bibfnamefont {David~P.}\
  \bibnamefont {Morgan}},\ }\href@noop {} {\emph {\bibinfo {title} {{Surface
  acoustic wave filters : with applications to electronic communications and
  signal processing}}}}\ (\bibinfo  {publisher} {Academic Press},\ \bibinfo
  {year} {2007})\ p.\ \bibinfo {pages} {429}\BibitemShut {NoStop}%
\bibitem [{\citenamefont {Peropadre}\ \emph {et~al.}(2013)\citenamefont
  {Peropadre}, \citenamefont {Lindkvist}, \citenamefont {Hoi}, \citenamefont
  {Wilson}, \citenamefont {Garcia-Ripoll}, \citenamefont {Delsing},\ and\
  \citenamefont {Johansson}}]{Peropadre2013}%
  \BibitemOpen
  \bibfield  {author} {\bibinfo {author} {\bibfnamefont {B}~\bibnamefont
  {Peropadre}}, \bibinfo {author} {\bibfnamefont {J}~\bibnamefont {Lindkvist}},
  \bibinfo {author} {\bibfnamefont {I-C}\ \bibnamefont {Hoi}}, \bibinfo
  {author} {\bibfnamefont {C~M}\ \bibnamefont {Wilson}}, \bibinfo {author}
  {\bibfnamefont {J~J}\ \bibnamefont {Garcia-Ripoll}}, \bibinfo {author}
  {\bibfnamefont {P}~\bibnamefont {Delsing}}, \ and\ \bibinfo {author}
  {\bibfnamefont {G}~\bibnamefont {Johansson}},\ }\bibfield  {title} {\enquote
  {\bibinfo {title} {{Scattering of coherent states on a single artificial
  atom}},}\ }\href {\doibase 10.1088/1367-2630/15/3/035009} {\bibfield
  {journal} {\bibinfo  {journal} {New Journal of Physics}\ }\textbf {\bibinfo
  {volume} {15}},\ \bibinfo {pages} {035009} (\bibinfo {year}
  {2013})}\BibitemShut {NoStop}%
\bibitem [{\citenamefont {Grimsmo}(2015)}]{Grimsmo2015}%
  \BibitemOpen
  \bibfield  {author} {\bibinfo {author} {\bibfnamefont {Arne~L.}\ \bibnamefont
  {Grimsmo}},\ }\bibfield  {title} {\enquote {\bibinfo {title} {{Time-Delayed
  Quantum Feedback Control}},}\ }\href {\doibase
  10.1103/PhysRevLett.115.060402} {\bibfield  {journal} {\bibinfo  {journal}
  {Physical Review Letters}\ }\textbf {\bibinfo {volume} {115}},\ \bibinfo
  {pages} {060402} (\bibinfo {year} {2015})}\BibitemShut {NoStop}%
\bibitem [{\citenamefont {Pichler}\ and\ \citenamefont
  {Zoller}(2016)}]{Pichler2016}%
  \BibitemOpen
  \bibfield  {author} {\bibinfo {author} {\bibfnamefont {Hannes}\ \bibnamefont
  {Pichler}}\ and\ \bibinfo {author} {\bibfnamefont {Peter}\ \bibnamefont
  {Zoller}},\ }\bibfield  {title} {\enquote {\bibinfo {title} {{Photonic
  Circuits with Time Delays and Quantum Feedback}},}\ }\href {\doibase
  10.1103/PhysRevLett.116.093601} {\bibfield  {journal} {\bibinfo  {journal}
  {Physical Review Letters}\ }\textbf {\bibinfo {volume} {116}},\ \bibinfo
  {pages} {093601} (\bibinfo {year} {2016})}\BibitemShut {NoStop}%
\bibitem [{\citenamefont {Forn-D{\'{i}}az}\ \emph {et~al.}(2018)\citenamefont
  {Forn-D{\'{i}}az}, \citenamefont {Lamata}, \citenamefont {Rico},
  \citenamefont {Kono},\ and\ \citenamefont {Solano}}]{Forn-Diaz2018}%
  \BibitemOpen
  \bibfield  {author} {\bibinfo {author} {\bibfnamefont {P.}~\bibnamefont
  {Forn-D{\'{i}}az}}, \bibinfo {author} {\bibfnamefont {L.}~\bibnamefont
  {Lamata}}, \bibinfo {author} {\bibfnamefont {E.}~\bibnamefont {Rico}},
  \bibinfo {author} {\bibfnamefont {J.}~\bibnamefont {Kono}}, \ and\ \bibinfo
  {author} {\bibfnamefont {E.}~\bibnamefont {Solano}},\ }\bibfield  {title}
  {\enquote {\bibinfo {title} {{Ultrastrong coupling regimes of light-matter
  interaction}},}\ }\href {http://arxiv.org/abs/1804.09275} {\  (\bibinfo
  {year} {2018})},\ \Eprint {http://arxiv.org/abs/1804.09275}
  {arXiv:1804.09275} \BibitemShut {NoStop}%
\bibitem [{\citenamefont {Kockum}\ \emph
  {et~al.}(2018{\natexlab{b}})\citenamefont {Kockum}, \citenamefont
  {Miranowicz}, \citenamefont {{De Liberato}}, \citenamefont {Savasta},\ and\
  \citenamefont {Nori}}]{Kockum2018}%
  \BibitemOpen
  \bibfield  {author} {\bibinfo {author} {\bibfnamefont {Anton~Frisk}\
  \bibnamefont {Kockum}}, \bibinfo {author} {\bibfnamefont {Adam}\ \bibnamefont
  {Miranowicz}}, \bibinfo {author} {\bibfnamefont {Simone}\ \bibnamefont {{De
  Liberato}}}, \bibinfo {author} {\bibfnamefont {Salvatore}\ \bibnamefont
  {Savasta}}, \ and\ \bibinfo {author} {\bibfnamefont {Franco}\ \bibnamefont
  {Nori}},\ }\bibfield  {title} {\enquote {\bibinfo {title} {{Ultrastrong
  coupling between light and matter}},}\ }\href
  {http://arxiv.org/abs/1807.11636} {\  (\bibinfo {year}
  {2018}{\natexlab{b}})},\ \Eprint {http://arxiv.org/abs/1807.11636}
  {arXiv:1807.11636} \BibitemShut {NoStop}%
\bibitem [{\citenamefont {Sanchez-Burillo}\ \emph {et~al.}(2014)\citenamefont
  {Sanchez-Burillo}, \citenamefont {Zueco}, \citenamefont {Garcia-Ripoll},\
  and\ \citenamefont {Martin-Moreno}}]{Sanchez-Burillo2014}%
  \BibitemOpen
  \bibfield  {author} {\bibinfo {author} {\bibfnamefont {E.}~\bibnamefont
  {Sanchez-Burillo}}, \bibinfo {author} {\bibfnamefont {D.}~\bibnamefont
  {Zueco}}, \bibinfo {author} {\bibfnamefont {J.~J.}\ \bibnamefont
  {Garcia-Ripoll}}, \ and\ \bibinfo {author} {\bibfnamefont {L.}~\bibnamefont
  {Martin-Moreno}},\ }\bibfield  {title} {\enquote {\bibinfo {title}
  {{Scattering in the Ultrastrong Regime: Nonlinear Optics with One Photon}},}\
  }\href {\doibase 10.1103/PhysRevLett.113.263604} {\bibfield  {journal}
  {\bibinfo  {journal} {Physical Review Letters}\ }\textbf {\bibinfo {volume}
  {113}},\ \bibinfo {pages} {263604} (\bibinfo {year} {2014})}\BibitemShut
  {NoStop}%
\bibitem [{\citenamefont {Nigg}\ \emph {et~al.}(2012)\citenamefont {Nigg},
  \citenamefont {Paik}, \citenamefont {Vlastakis}, \citenamefont {Kirchmair},
  \citenamefont {Shankar}, \citenamefont {Frunzio}, \citenamefont {Devoret},
  \citenamefont {Schoelkopf},\ and\ \citenamefont {Girvin}}]{Nigg2012}%
  \BibitemOpen
  \bibfield  {author} {\bibinfo {author} {\bibfnamefont {Simon~E.}\
  \bibnamefont {Nigg}}, \bibinfo {author} {\bibfnamefont {Hanhee}\ \bibnamefont
  {Paik}}, \bibinfo {author} {\bibfnamefont {Brian}\ \bibnamefont {Vlastakis}},
  \bibinfo {author} {\bibfnamefont {Gerhard}\ \bibnamefont {Kirchmair}},
  \bibinfo {author} {\bibfnamefont {S.}~\bibnamefont {Shankar}}, \bibinfo
  {author} {\bibfnamefont {Luigi}\ \bibnamefont {Frunzio}}, \bibinfo {author}
  {\bibfnamefont {M.~H.}\ \bibnamefont {Devoret}}, \bibinfo {author}
  {\bibfnamefont {R.~J.}\ \bibnamefont {Schoelkopf}}, \ and\ \bibinfo {author}
  {\bibfnamefont {S.~M.}\ \bibnamefont {Girvin}},\ }\bibfield  {title}
  {\enquote {\bibinfo {title} {{Black-Box Superconducting Circuit
  Quantization}},}\ }\href {\doibase 10.1103/PhysRevLett.108.240502} {\bibfield
   {journal} {\bibinfo  {journal} {Physical Review Letters}\ }\textbf {\bibinfo
  {volume} {108}},\ \bibinfo {pages} {240502} (\bibinfo {year}
  {2012})}\BibitemShut {NoStop}%
\bibitem [{\citenamefont {Forn-D{\'{i}}az}\ \emph {et~al.}(2017)\citenamefont
  {Forn-D{\'{i}}az}, \citenamefont {Garc{\'{i}}a-Ripoll}, \citenamefont
  {Peropadre}, \citenamefont {Orgiazzi}, \citenamefont {Yurtalan},
  \citenamefont {Belyansky}, \citenamefont {Wilson},\ and\ \citenamefont
  {Lupascu}}]{Forn-Diaz2016}%
  \BibitemOpen
  \bibfield  {author} {\bibinfo {author} {\bibfnamefont {P.}~\bibnamefont
  {Forn-D{\'{i}}az}}, \bibinfo {author} {\bibfnamefont {J.~J.}\ \bibnamefont
  {Garc{\'{i}}a-Ripoll}}, \bibinfo {author} {\bibfnamefont {B.}~\bibnamefont
  {Peropadre}}, \bibinfo {author} {\bibfnamefont {J.~L.}\ \bibnamefont
  {Orgiazzi}}, \bibinfo {author} {\bibfnamefont {M.~A.}\ \bibnamefont
  {Yurtalan}}, \bibinfo {author} {\bibfnamefont {R.}~\bibnamefont {Belyansky}},
  \bibinfo {author} {\bibfnamefont {C.~M.}\ \bibnamefont {Wilson}}, \ and\
  \bibinfo {author} {\bibfnamefont {A.}~\bibnamefont {Lupascu}},\ }\bibfield
  {title} {\enquote {\bibinfo {title} {{Ultrastrong coupling of a single
  artificial atom to an electromagnetic continuum in the nonperturbative
  regime}},}\ }\href {\doibase 10.1038/nphys3905} {\bibfield  {journal}
  {\bibinfo  {journal} {Nature Physics}\ }\textbf {\bibinfo {volume} {13}},\
  \bibinfo {pages} {39--43} (\bibinfo {year} {2017})},\ \Eprint
  {http://arxiv.org/abs/1602.00416} {arXiv:1602.00416} \BibitemShut {NoStop}%
\bibitem [{\citenamefont {D{\'{i}}az-Camacho}\ \emph
  {et~al.}(2016)\citenamefont {D{\'{i}}az-Camacho}, \citenamefont {Bermudez},\
  and\ \citenamefont {Garc{\'{i}}a-Ripoll}}]{Diaz-Camacho2016}%
  \BibitemOpen
  \bibfield  {author} {\bibinfo {author} {\bibfnamefont {Guillermo}\
  \bibnamefont {D{\'{i}}az-Camacho}}, \bibinfo {author} {\bibfnamefont
  {Alejandro}\ \bibnamefont {Bermudez}}, \ and\ \bibinfo {author}
  {\bibfnamefont {Juan~Jos{\'{e}}}\ \bibnamefont {Garc{\'{i}}a-Ripoll}},\
  }\bibfield  {title} {\enquote {\bibinfo {title} {{Dynamical polaron Ansatz :
  A theoretical tool for the ultrastrong-coupling regime of circuit QED}},}\
  }\href {\doibase 10.1103/PhysRevA.93.043843} {\bibfield  {journal} {\bibinfo
  {journal} {Physical Review A}\ }\textbf {\bibinfo {volume} {93}},\ \bibinfo
  {pages} {043843} (\bibinfo {year} {2016})},\ \Eprint
  {http://arxiv.org/abs/1512.04244} {arXiv:1512.04244} \BibitemShut {NoStop}%
\bibitem [{\citenamefont {Krimer}\ \emph {et~al.}(2014)\citenamefont {Krimer},
  \citenamefont {Liertzer}, \citenamefont {Rotter},\ and\ \citenamefont
  {T{\"{u}}reci}}]{Krimer2014}%
  \BibitemOpen
  \bibfield  {author} {\bibinfo {author} {\bibfnamefont {Dmitry~O.}\
  \bibnamefont {Krimer}}, \bibinfo {author} {\bibfnamefont {Matthias}\
  \bibnamefont {Liertzer}}, \bibinfo {author} {\bibfnamefont {Stefan}\
  \bibnamefont {Rotter}}, \ and\ \bibinfo {author} {\bibfnamefont {Hakan~E.}\
  \bibnamefont {T{\"{u}}reci}},\ }\bibfield  {title} {\enquote {\bibinfo
  {title} {{Route from spontaneous decay to complex multimode dynamics in
  cavity QED}},}\ }\href {\doibase 10.1103/PhysRevA.89.033820} {\bibfield
  {journal} {\bibinfo  {journal} {Physical Review A}\ }\textbf {\bibinfo
  {volume} {89}},\ \bibinfo {pages} {033820} (\bibinfo {year}
  {2014})}\BibitemShut {NoStop}%
\bibitem [{\citenamefont {Guo}\ \emph {et~al.}(2017)\citenamefont {Guo},
  \citenamefont {Grimsmo}, \citenamefont {Kockum}, \citenamefont {Pletyukhov},\
  and\ \citenamefont {Johansson}}]{Guo2017}%
  \BibitemOpen
  \bibfield  {author} {\bibinfo {author} {\bibfnamefont {Lingzhen}\
  \bibnamefont {Guo}}, \bibinfo {author} {\bibfnamefont {Arne}\ \bibnamefont
  {Grimsmo}}, \bibinfo {author} {\bibfnamefont {Anton~Frisk}\ \bibnamefont
  {Kockum}}, \bibinfo {author} {\bibfnamefont {Mikhail}\ \bibnamefont
  {Pletyukhov}}, \ and\ \bibinfo {author} {\bibfnamefont {G{\"{o}}ran}\
  \bibnamefont {Johansson}},\ }\bibfield  {title} {\enquote {\bibinfo {title}
  {{Giant acoustic atom: A single quantum system with a deterministic time
  delay}},}\ }\href {\doibase 10.1103/PhysRevA.95.053821} {\bibfield  {journal}
  {\bibinfo  {journal} {Physical Review A}\ }\textbf {\bibinfo {volume} {95}},\
  \bibinfo {pages} {053821} (\bibinfo {year} {2017})}\BibitemShut {NoStop}%
\end{thebibliography}%

\end{document}